\documentclass[12pt,english,floatfix,superscriptaddress,aps,prd,preprint]{revtex4}
\usepackage[utf8]{inputenc}
\usepackage{float}
\usepackage{array}
\usepackage{lipsum}
\usepackage{graphicx}
\usepackage{amsmath,amsthm,amsfonts,amssymb}
\usepackage{graphicx}
\usepackage[english]{babel}
\usepackage{color}
\usepackage{tensor}
\usepackage{esint}
\usepackage[dvips]{epsfig}
\usepackage[dvips]{graphicx}
\usepackage{float}
\usepackage{units}
\usepackage{textcomp}
\usepackage{mathrsfs}
\usepackage{amsmath}
\usepackage[makeroom]{cancel}
\usepackage{amssymb}
\usepackage{amsbsy}
\usepackage{amsfonts}
\usepackage{amssymb,mathrsfs,xcolor}
\usepackage{esint}
\usepackage{array}
\usepackage{graphicx}

\usepackage{hyperref}
\hypersetup{
    colorlinks,
    citecolor=blue,
    filecolor=green,
    linkcolor=purple,
    urlcolor=red,
}

\usepackage{slashed}

\newcommand{\ie}{\begin{equation}}
\newcommand{\fe}{\end{equation}}
\newcommand{\se}{\begin{eqnarray}}
\newcommand{\ff}{\end{eqnarray}}

\usepackage{hyperref}
\hypersetup{colorlinks,breaklinks,
			citecolor=[rgb]{0,0.0,1.0},
            urlcolor=[rgb]{0.0,0.0,1.0},
            linkcolor=[rgb]{0,0.5,0.9}}

\begin{document}

\title{Higher-derivative Lorentz-breaking dispersion relations: a thermal description}

\author{A. A. Ara\'{u}jo Filho}
\email{dilto@fisica.ufc.br}

\affiliation{Universidade Federal do Cear\'a (UFC), Departamento de F\'isica,\\ Campus do Pici,
Fortaleza - CE, C.P. 6030, 60455-760 - Brazil.}

\author{A. Yu. Petrov}
\email{petrov@fisica.ufpb.br}
\affiliation{Departamento de Física, Universidade Federal da Paraíba, Caixa Postal 5008, 58051-970, João Pessoa, Paraíba,  Brazil.}




\date{\today}

\begin{abstract}
This paper is devoted to study the thermal aspects of a photon gas within the context of Planck-scale-modified dispersion relations. We study the spectrum of radiation and the correction to the \textit{Stefan–Boltzmann} law in different cases when the Lorentz symmetry is no longer preserved. Explicitly, we examine two models within the context of CPT-even and CPT-odd sectors respectively. To do so, three distinct scenarios of the Universe are considered: the Cosmic Microwave Background (CMB), the electroweak epoch, and the inflationary era. Moreover, the equations of state in these cases turn out to display a dependence on Lorentz-breaking parameters. Finally, we also provide for both theories the analyses of the Helmholtz free energy, the mean energy, the entropy and the heat capacity.

\end{abstract}


\maketitle

\section{Introduction}

Although the Lorentz invariance is a well-established symmetry of the nature, its possible violation is assumed within many contexts \cite{liberati2014astrophysical,tasson2014we,hees2016tests}. During the last years, the Lorentz symmetry breaking and its possible implications are intensively studied in different scenarios, see f.e. \cite{maluf2019antisymmetric,alfaro2002loop,pospelov2012lorentz,mavromatos2007lorentz,carroll2001noncommutative}. As it is known, the Lorentz-breaking parameters are experimentally presumed to be very tiny \cite{kostelecky2011data}. Nevertheless, this does not imply that they contribute to the physical processes in a negligible manner. Moreover, the presence of Lorentz-violating (LV) higher-derivative additive terms can imply in arising of large quantum corrections \cite{collins2004lorentz,nascimento2018renormalization}. 

All this clearly motivates us to investigate the higher-derivative LV terms. The first known example of such term was originally introduced by Myers and Pospelov many years ago \cite{myers2003ultraviolet}. The key feature of this term is that it involves higher (third) derivatives, being proportional to a Lorentz-breaking constant vector and to a small constant parameter (treated as an inverse of some large mass scale). Furthermore, various issues related to them were studied, including its perturbative generation and dispersion relations (for a review on such terms see f.e. \cite{kostelecky2009electrodynamics}. An exhaustive list of such terms with dimensions up to 6 can be found in \cite{kostelecky2019gauge}). It was argued in Ref. \cite{mariz2019quantum} that a theory involving such a term can be treated consistently as a series in the above-mentioned inverse mass scale.

One of the important issues related to higher-derivative LV theories is certainly their thermodynamical aspects. Studies on the thermodynamics of LV extensions of various field theory models could provide additional information about initial stages of expansion of the Universe, whose size at these stages was compatible with characteristic scales of Lorentz symmetry breaking \cite{kostelecky2011data}. The methodology for studying the thermodynamical aspects in LV theories has firstly been presented in \cite{colladay2004statistical}. Since then, various applications of this procedure have been developed \cite{casana2008lorentz,casana2009finite,gomes2010free,aa2020lorentz, maluf2020thermodynamic,reis2020thermal,anacleto2018lorentz,das2009relativistic,petrov2021bouncing,reis2021fermions}. However, the higher-derivative cases have not been much explored up to now in this context.

In this paper, we follow the procedure proposed by Amelino-Camelia \cite{amelino2001testable} where the starting point in the study is the LV dispersion relations rather than the Lagrangian formalism of the corresponding field models. Nevertheless, we note that in many cases it is possible to indicate at least some simplified models yielding such dispersion relations.  We note that the dispersion relations that we consider in the present manuscript might possibly be used to describe some quantum gravity effects (see discussions in \cite{amelino2001testable}), or, at least, they can probably arise in some LV extensions of QED. We consider two examples, CPT-even and CPT-odd ones. In principle, our results involving thermal radiation may be confronted with experimental data as soon as they are available. In this sense, our study might help in the investigation of any trace of the Lorentz violation within cosmological scenarios concerning thermal radiation as the starting point.


\section{Thermodynamical aspects of CPT-even higher-derivative LV theory}
 
Here, let us consider the higher-derivative LV theories. To study their thermodynamical aspects, we will define the dispersion relations of these theories in which are generally sufficient to obtain various related quantities such as free energy \cite{gomes2010free}. Our first example is the following
dispersion relation \cite{bianco2018modified}:
\ie
E^{2} = {\bf{k}}^{2} + \sigma^{4} {\bf{k}}^{6}. \label{d1}
\fe
Clearly, if we consider $\sigma \rightarrow 0$, the usual dispersion relation $E^{2}={\bf{k}}^{2}$ is recove. Physically, such a relation can arise in Horava-Lifshitz-like theories involving both $z=1$ and $z=3$ terms in the spatial sector, e.g. in a scalar field model characterized by the Lagrangian ${\cal L}=\frac{1}{2}\phi(\Box+\sigma^4(\nabla^2)^3)\phi$, in a spinor model involving terms with these values of $z$ \cite{gomes2018one}, and perhaps for some degrees of freedom of a certain LV gauge theory or, a Horava-Lifshitz-like gravity model involving $z=3$ and $z=1$ terms \cite{Horava:2009uw}.
Let us briefly discuss the possible physical significance of these relations. It must be noted that the usual Lorentz-breaking extensions of various field theory models conside within phernomenological context -- their full list is given in \cite{kostelecky2011data} -- have rather a different form since they involve the same orders in space and time derivatives until we choose the special form of Lorentz-breaking parameters. More so, the models involving the dispersion relations like (\ref{d1}) can also be regarded for studies of various physical problems; for instance, one of the most important applications of this model has been developed in \cite{deFarias:2020iqh} where it was used for an investigation of gamma-ray bursts in the Lorentz-breaking context
and allowed to estimate the characteristic energy of quantum gravity mass.
It is easy to see that the dispersion relation in Eq. (\ref{d1})
gives rise to six solutions. Nevertheless, only one of them allows us to work on a positive defined real spatial momentum. In this sense, we can write the solution of Eq. (\ref{d1}) as being
\ie
{\bf{k}} = \left(\frac{\sqrt[3]{\sqrt{3} \sqrt{4 \sigma ^{12}+27 \sigma^{16} E^4}+9 \sigma^8 E^2}}{\sqrt[3]{2} \, 3^{2/3} \sigma ^4}-\frac{\sqrt[3]{\frac{2}{3}}}{\sqrt[3]{\sqrt{3} \sqrt{4 \sigma ^{12}+27 \sigma ^{16} E^4}+9 \sigma ^8 E^2}}\right)^{1/2}. \label{dr1}
\fe
Next, we take the advantage of using the formalism of the partition function in order to derive all relevant thermodynamic quantities i.e., Helmholtz free energy, mean energy, entropy and heat capacity. Initially, we calculate the number of accessible states of the system \cite{reif1998fundamentals,gibbs2014elementary,greiner2012thermodynamics,callen1998thermodynamics}. By definition, it can be represented as
\ie
\Omega(E) = \frac{\zeta}{(2\pi)^{3}}\int \int \mathrm{d}^{3} {\bf{x}} \,\mathrm{d}^{3} {\bf{k}}, \label{preacessible}
\fe
where $\zeta$ is the spin multiplicity whose magnitude in the photon sector is $\zeta = 2$ \cite{anacleto2018lorentz}. More so, Eq. (\ref{preacessible}) can be rewritten as
\ie
\Omega(E) = \frac{V}{\pi^{2}} \int^{\infty}_{0} \mathrm{d} {\bf{k}} |{\bf{k}}|^{2}, \label{ms1}
\fe
where $V$ is the volume of the thermal reservoir and the integral measure $\mathrm{d} {\bf{k}}$ is given by
\ie
\mathrm{d} {\bf{k}} = \frac{\frac{\sqrt[3]{\frac{2}{3}} \left(\frac{54 \sqrt{3} \sigma ^{16} E^3}{\sqrt{4 \sigma ^{12}+27 \sigma ^{16} E^4}}+18 \sigma^8 E\right)}{3 \left(\sqrt{3} \sqrt{4 \sigma^{12}+27 \sigma^{16} E^4}+9 \sigma^8 E^2\right)^{4/3}}+\frac{\frac{54 \sqrt{3} \sigma^{16} E^3}{\sqrt{4 \sigma^{12}+27 \sigma^{16} E^4}}+18 \sigma^8 E}{3 \sqrt[3]{2} 3^{2/3} \sigma^4 \left(\sqrt{3} \sqrt{4 \sigma^{12}+27 \sigma^{16} E^4}+9 \sigma^ 8 E^2\right)^{2/3}}}{2 \sqrt{\frac{\sqrt[3]{\sqrt{3} \sqrt{4 \sigma^{12}+27 \sigma^{16} E^4}+9 \sigma^8 E^2}}{\sqrt[3]{2} 3^{2/3} \sigma^4}-\frac{\sqrt[3]{\frac{2}{3}}}{\sqrt[3]{\sqrt{3} \sqrt{4 \sigma^{12}+27 \sigma^{16} E^4}+9 \sigma^8 E^2}}}} \mathrm{d}E.
\label{vol1}
\fe
Next, we substitute $(\ref{dr1})$ and $(\ref{vol1})$ in $(\ref{ms1})$, which yields
\ie
\begin{split}
\Omega(\sigma) = \frac{V}{\pi^{2}} \int^{\infty}_{0}
& \left(\frac{\sqrt[3]{\frac{2}{3}} \left(\frac{54 \sqrt{3} \sigma^{16} E^3}{\sqrt{4 \sigma ^{12}+27 \sigma^{16} E^4}} + 18 \sigma^8 E\right)}{3 \left(\sqrt{3} \sqrt{4 \sigma ^{12}+27 \sigma^{16} E^4}+9 \sigma^8 E^2\right)^{4/3}} \right.\\
&\left. +
\frac{\frac{54 \sqrt{3} \sigma^{16} E^3}{\sqrt{4 \sigma^{12}+27 \sigma^{16} E^4}}+18 \sigma^8 E}{3 \sqrt[3]{2} 3^{2/3} \sigma^4 \left(\sqrt{3} \sqrt{4 \sigma^{12}+27 \sigma^{16} E^4}+9 \sigma^8 E^2\right)^{2/3}}\right) \\
& \times \sqrt{\frac{\sqrt[3]{\sqrt{3} \sqrt{4 \sigma ^{12}+27 \sigma^{16} E^4}+9 \sigma^8 E^2}}{\sqrt[3]{2} 3^{2/3} \sigma^4}-\frac{\sqrt[3]{\frac{2}{3}}}{\sqrt[3]{\sqrt{3} \sqrt{4 \sigma^{12}+27 \sigma^{16} E^4}+9 \sigma^8 E^2}}}\,\mathrm{d}E,
\end{split}
\fe 
and, therefore, we can explicitly write down  the partition function in a manner similar to Refs. \cite{anacleto2018lorentz,maluf2020casimir} as follows:
\ie
\begin{split}
\mathrm{ln}\left[ Z(\beta,\Gamma,\sigma)\right] = &- \frac{V}{\pi^{2}}  \int^{\infty}_{0}
 \left(\frac{\sqrt[3]{\frac{2}{3}} \left(\frac{54 \sqrt{3} \sigma^{16} E^3}{\sqrt{4 \sigma ^{12}+27 \sigma^{16} E^4}} + 18 \sigma^8 E\right)}{3 \left(\sqrt{3} \sqrt{4 \sigma ^{12}+27 \sigma^{16} E^4}+9 \sigma^8 E^2\right)^{4/3}} \right.\\
&\left. +
\frac{\frac{54 \sqrt{3} \sigma^{16} E^3}{\sqrt{4 \sigma^{12}+27 \sigma^{16} E^4}}+18 \sigma^8 E}{3 \sqrt[3]{2} 3^{2/3} \sigma^4 \left(\sqrt{3} \sqrt{4 \sigma^{12}+27 \sigma^{16} E^4}+9 \sigma^8 E^2\right)^{2/3}}\right) \times \mathrm{ln} \left(  1- e^{-\beta E} \right) \\
& \times \sqrt{\frac{\sqrt[3]{\sqrt{3} \sqrt{4 \sigma ^{12}+27 \sigma^{16} E^4}+9 \sigma^8 E^2}}{\sqrt[3]{2} 3^{2/3} \sigma^4}-\frac{\sqrt[3]{\frac{2}{3}}}{\sqrt[3]{\sqrt{3} \sqrt{4 \sigma^{12}+27 \sigma^{16} E^4}+9 \sigma^8 E^2}}}\,\mathrm{d}E, \label{partitioncase1}
\end{split}
\fe 
where $\beta = 1/(k_{B}T)$ is the inverse of the temperature. From Eq. (\ref{partitioncase1}), the thermodynamic functions can be derived. It is important to mention that all our calculations will provide
the values of these functions per volume $V$. The thermal functions of interest are defined as
\ie
\begin{split}
 & F(\beta,\sigma)=-\frac{1}{\beta} \mathrm{ln}\left[Z(\beta,\sigma)\right], \\
 & U(\beta,\sigma)=-\frac{\partial}{\partial\beta} \mathrm{ln}\left[Z(\beta,\sigma)\right], \\
 & S(\beta,\sigma)=k_B\beta^2\frac{\partial}{\partial\beta}F(\beta,\sigma), \\
 & C_V(\beta,\sigma)=-k_B\beta^2\frac{\partial}{\partial\beta}U(\beta,\sigma).
\label{properties}
\end{split}
\fe  
Primarily, let us focus on the mean energy
\begin{eqnarray}
U(\beta,\sigma) &=& \frac{V}{\pi^{2}}  \int^{\infty}_{0} \mathrm{d}E E \left(\frac{\sqrt[3]{\frac{2}{3}} \left(\frac{54 \sqrt{3} \sigma^{16} E^3}{\sqrt{4 \sigma ^{12}+27 \sigma^{16} E^4}} + 18 \sigma^8 E\right)}{3 \left(\sqrt{3} \sqrt{4 \sigma ^{12}+27 \sigma^{16} E^4}+9 \sigma^8 E^2\right)^{4/3}} \right.\nonumber\\
&+&\left. 
\frac{\frac{54 \sqrt{3} \sigma^{16} E^3}{\sqrt{4 \sigma^{12}+27 \sigma^{16} E^4}}+18 \sigma^8 E}{3 \sqrt[3]{2} 3^{2/3} \sigma^4 \left(\sqrt{3} \sqrt{4 \sigma^{12}+27 \sigma^{16} E^4}+9 \sigma^8 E^2\right)^{2/3}}\right) \times \frac{e^{-\beta E}} {\left(  1- e^{-\beta E} \right)}\nonumber\\
& \times& \sqrt{\frac{\sqrt[3]{\sqrt{3} \sqrt{4 \sigma ^{12}+27 \sigma^{16} E^4}+9 \sigma^8 E^2}}{\sqrt[3]{2} \cdot 3^{2/3} \sigma^4}-\frac{\sqrt[3]{\frac{2}{3}}}{\sqrt[3]{\sqrt{3} \sqrt{4 \sigma^{12}+27 \sigma^{16} E^4}+9 \sigma^8 E^2}}}\,\label{meanenergy1}
\end{eqnarray}
which implies the following spectral radiance:
\begin{eqnarray}
\chi(\sigma,\nu) &=& \frac{h \nu}{\pi^{2}} \left(\frac{\sqrt[3]{\frac{2}{3}} \left(\frac{54 \sqrt{3} \sigma^{16} (h\nu)^3}{\sqrt{4 \sigma^{12}+27 \sigma^{16} (h\nu)^4}} + 18 \sigma^8 (h\nu)\right)}{3 \left(\sqrt{3} \sqrt{4 \sigma ^{12}+27 \sigma^{16} (h\nu)^4}+9 \sigma^8 (h\nu)^2\right)^{4/3}} \right.\label{meanenergy1a}\\
&+&\left. 
\frac{\frac{54 \sqrt{3} \sigma^{16} (h\nu)^3}{\sqrt{4 \sigma^{12}+27 \sigma^{16} (h\nu)^4}}+18 \sigma^8 (h\nu)}{3 \sqrt[3]{2} 3^{2/3} \sigma^4 \left(\sqrt{3} \sqrt{4 \sigma^{12}+27 \sigma^{16} (h\nu)^4}+9 \sigma^8 (h\nu)^2\right)^{2/3}}\right) \times \frac{e^{-\beta (h\nu)}} {\left(  1- e^{-\beta (h\nu)} \right)\nonumber}\\
&\times&  \sqrt{\frac{\sqrt[3]{\sqrt{3} \sqrt{4 \sigma ^{12}+27 \sigma^{16} (h\nu)^4}+9 \sigma^8 (h\nu)^2}}{\sqrt[3]{2} 3^{2/3} \sigma^4}-\frac{\sqrt[3]{\frac{2}{3}}}{\sqrt[3]{\sqrt{3} \sqrt{4 \sigma^{12}+27 \sigma^{16} (h\nu)^4}+9 \sigma^8 (h\nu)^2}}}\,.\nonumber
\end{eqnarray}
Here, $E=h\nu$ where $h$ is the well-known Planck constant and $\nu$ is the frequency. Now, let us examine how the parameter $\sigma$ affects the spectral radiance of our theory. In addition, it must be noted that, despite of showing explicitly the constants $h,k_{B}$, to obtain the following calculations, we choose them as being $h=k_{B}=1$. At the beginning, we examine how the black body spectra can be modified by the parameter $\sigma$. Moreover, we consider three different configurations of temperatures for our system, namely, CMB ($\mathrm{T}= 10^{-13}$ GeV), electroweak epoch ($\mathrm{T}= 10^{3}$ GeV) and inflationary era ($\mathrm{T}= 10^{13}$ GeV). 

\begin{figure}[ht]
\centering
\includegraphics[width=8cm,height=5cm]{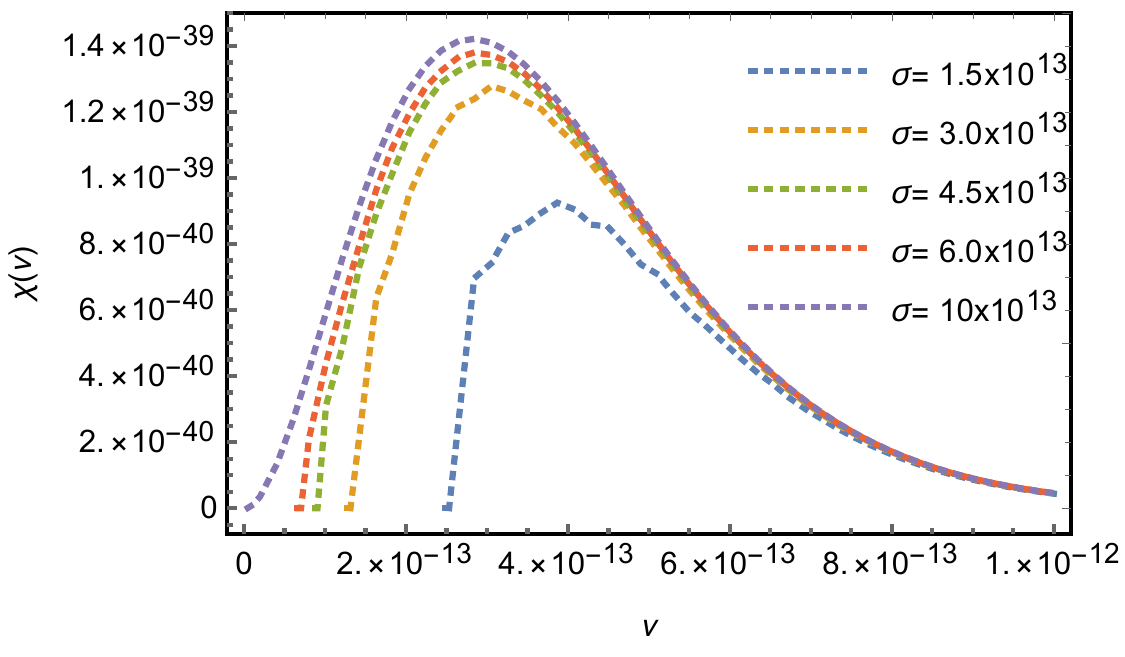}
\includegraphics[width=8cm,height=5cm]{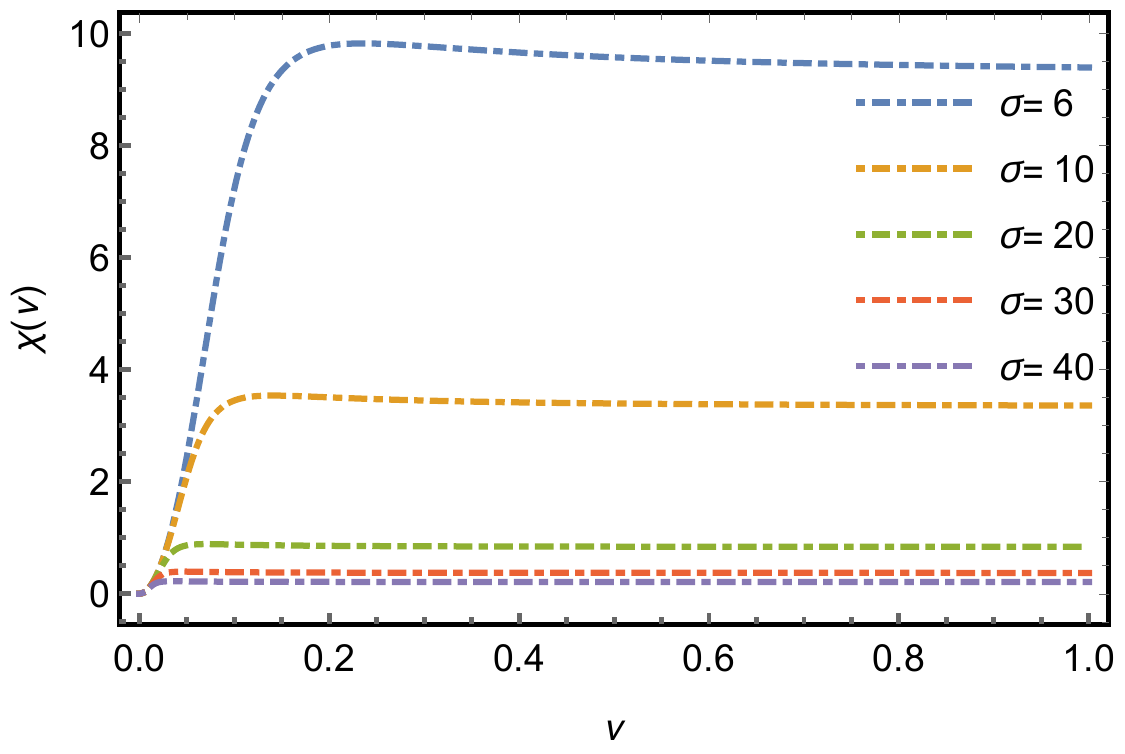}
\includegraphics[width=8cm,height=5cm]{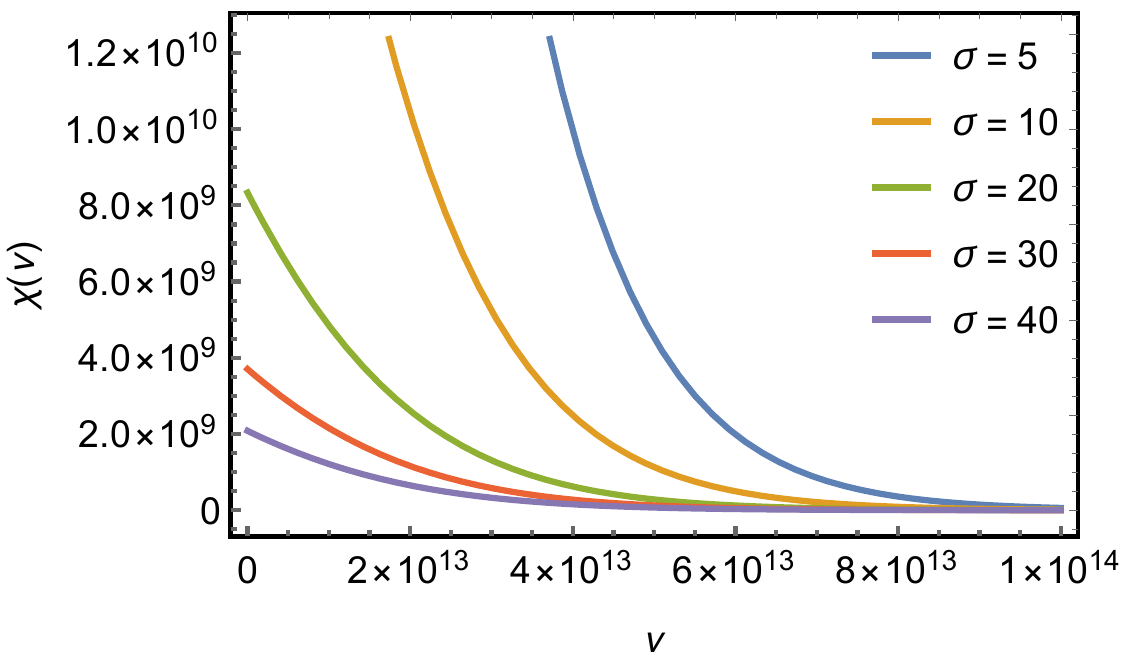}
\caption{The plots exhibit the spectral radiance $\chi(\nu)$ changing for different values of frequency $\nu$ and the Lorentz-breaking parameter $\sigma$ (its unit is GeV$^{-1}$). The top left (dotted) is the configuration to the cosmic microwave background, i.e., $\beta = 10^{13}$ GeV$^{-1}$; the top right (dot-dashed) is ascribed to the electroweak configuration, i.e., $\beta = 10^{-3}$ GeV$^{-1}$; the bottom plot shows the black body radiation to the inflationary period of the Universe, i.e., $\beta = 10^{-13}$ GeV$^{-1}$.  }
\label{spectral-radiances1}
\end{figure}

The results displayed at Fig. \ref{spectral-radiances1} show that the only one configuration of the black body radiation took a proper place in a prominent manner (in terms of shape of the curve) -- it was with the CMB temperature. Furthermore, taking into account the electroweak scenario, we see that the graphics started to increase reaching maxima peaks and, then, tended to attenuate their values until having a constant behavior. On the other hand, to the inflationary temperature, the plots show a behavior closer to Wien’s energy density distribution \cite{zettili2003quantum}.

Another interesting aspect to be verified is the correction to the \textit{Stefan–Boltzmann} law ascribed to the parameter $\sigma$. In order to do this, let us define the constant: 
\ie
\tilde{\alpha} \equiv U(\beta,\sigma) \beta^{4}. \label{sbla}
\fe

\begin{figure}[ht]
\centering
\includegraphics[width=8cm,height=5cm]{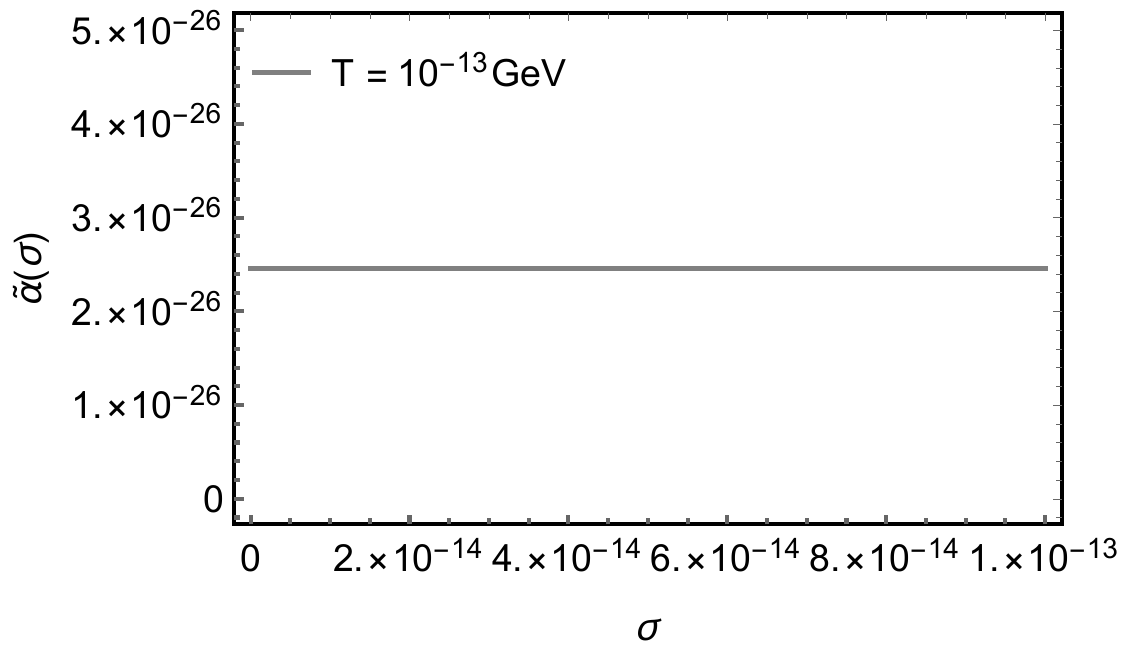}
\includegraphics[width=8cm,height=5cm]{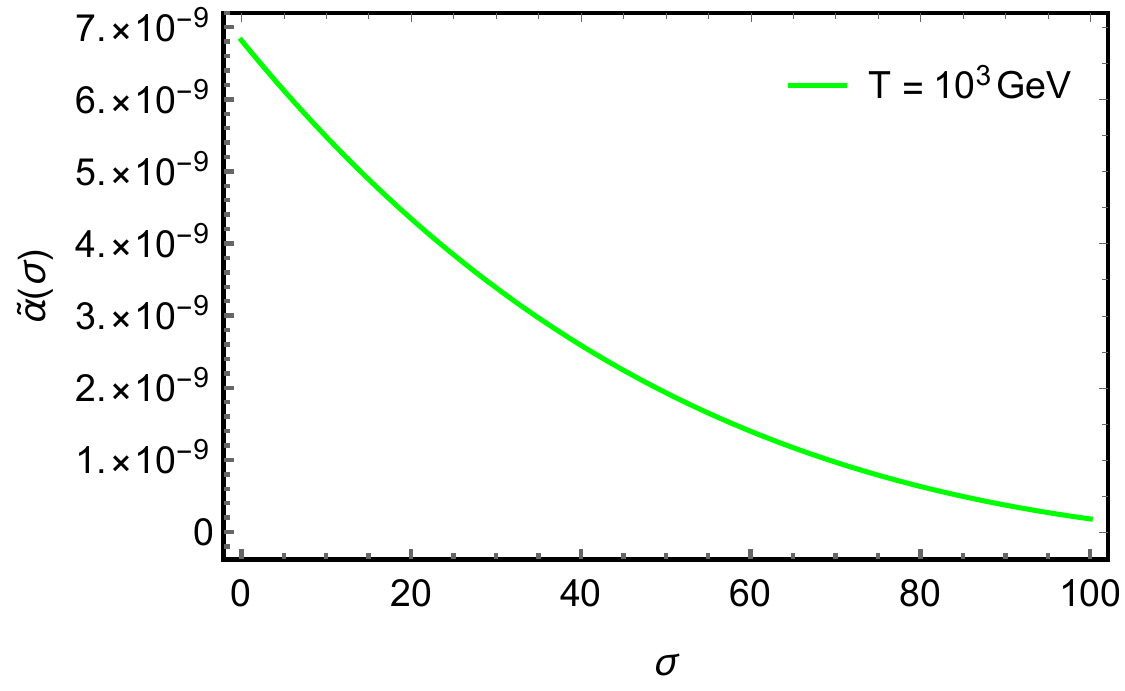}
\includegraphics[width=8cm,height=5cm]{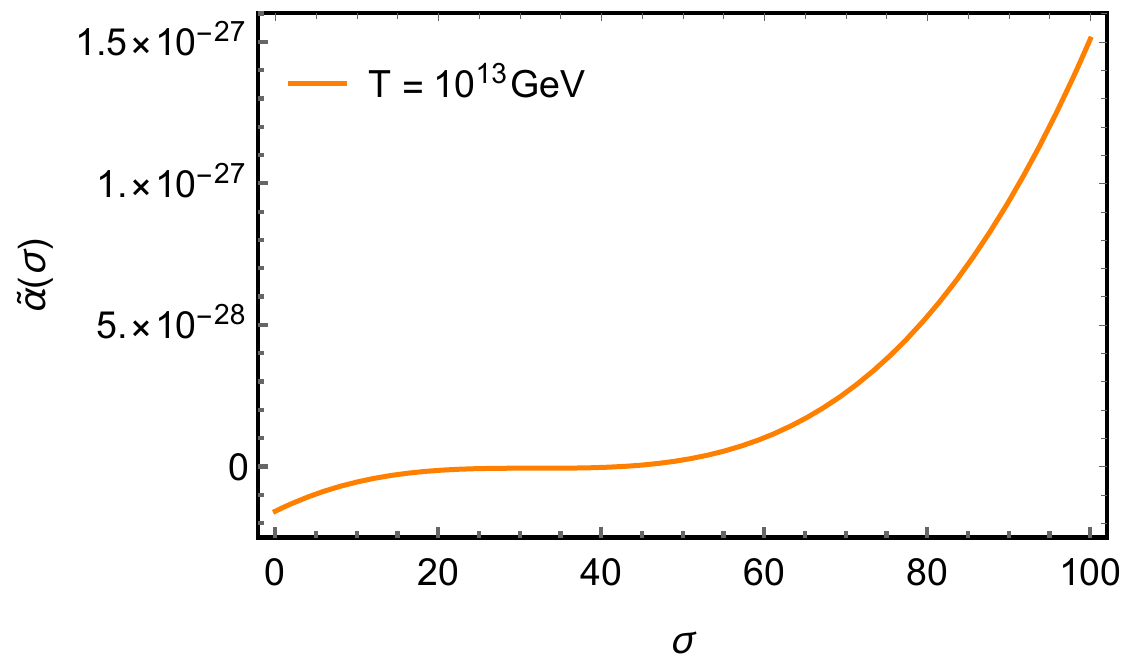}
\caption{The figure shows the correction to the \textit{Stefan–Boltzmann} law ascribed to parameter $\tilde{\alpha}$ as a function of $\sigma$ (its unit is GeV$^{-1}$) for the temperatures of cosmic microwave background (top left), electroweak scenario (top right) and the early inflationary universe (bottom).}
\label{alphas1}
\end{figure}

As it can easily be noted, the above expressions are unsolvable analytically. Thereby, the calculations will be performed numerically. Analogously, we regard the same previous configurations of temperature and the plots are exhibited in Fig. \ref{alphas1}. For the CMB temperature, the curve exhibits a constant behavior of $\alpha$. Moreover, to the electroweak epoch, we have a decreasing function $\tilde{\alpha}$ when $\sigma$ started to increase. The inflationary era, on the other hand, shows an increasing function of $\tilde{\alpha}$ for positive changes of $\sigma$. To this latter case, for $\tilde{\alpha} < 20$, the system seems to show instability. Next, we shall acquire all the remaining thermodynamic properties in what follows.

\begin{figure}[ht]
\centering
\includegraphics[width=8cm,height=5cm]{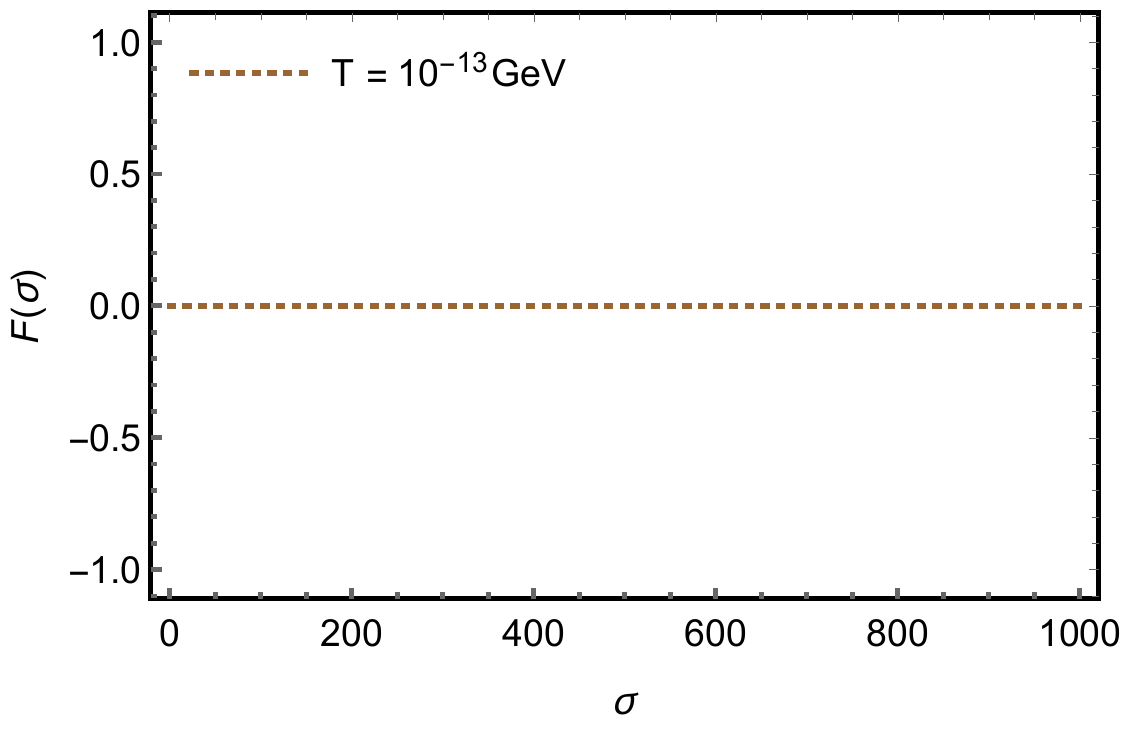}
\includegraphics[width=8cm,height=5cm]{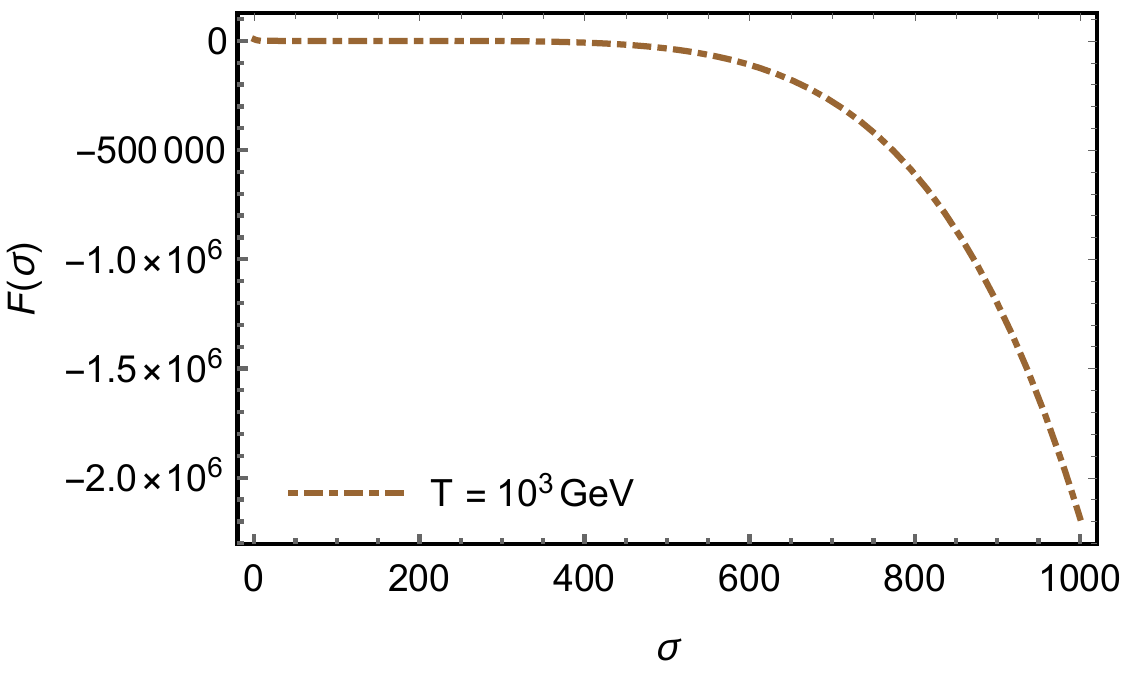}
\includegraphics[width=8cm,height=5cm]{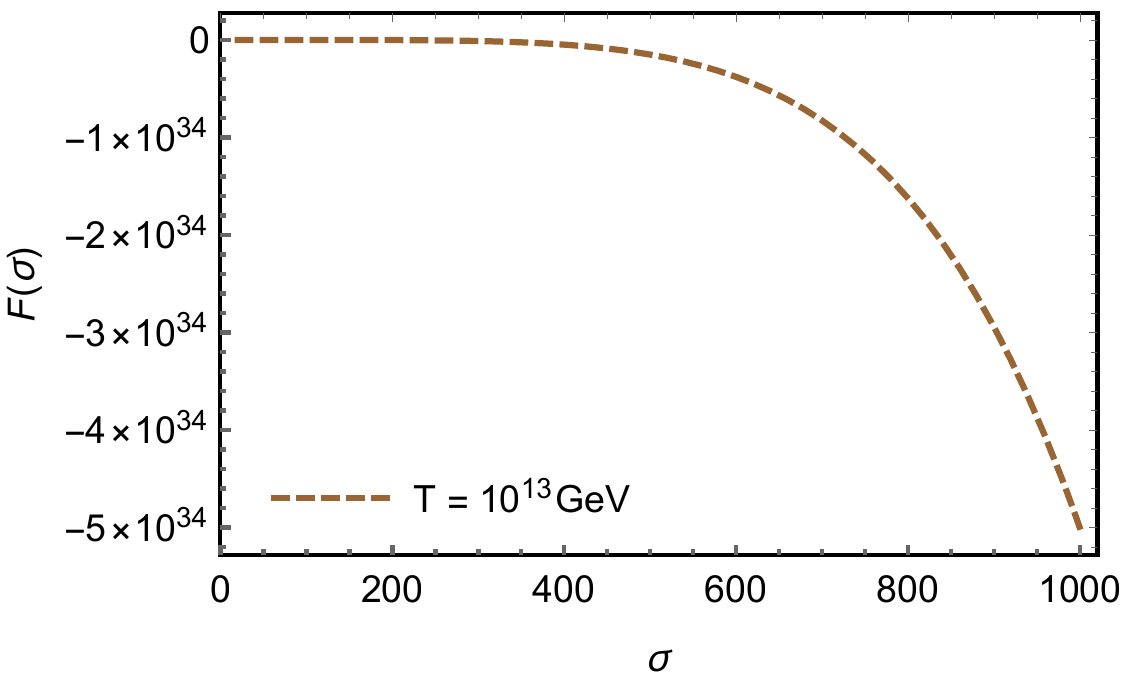}
\caption{The figure shows the modification of the Helmholtz free energy $F(\sigma)$ due to the parameter $\sigma$ (its unit is GeV$^{-1}$) considering the temperatures of cosmic microwave background (top left), electroweak scenario (top right) and the early inflationary universe (bottom).}
\label{fs1}
\end{figure}

\begin{figure}[ht]
\centering
\includegraphics[width=8cm,height=5cm]{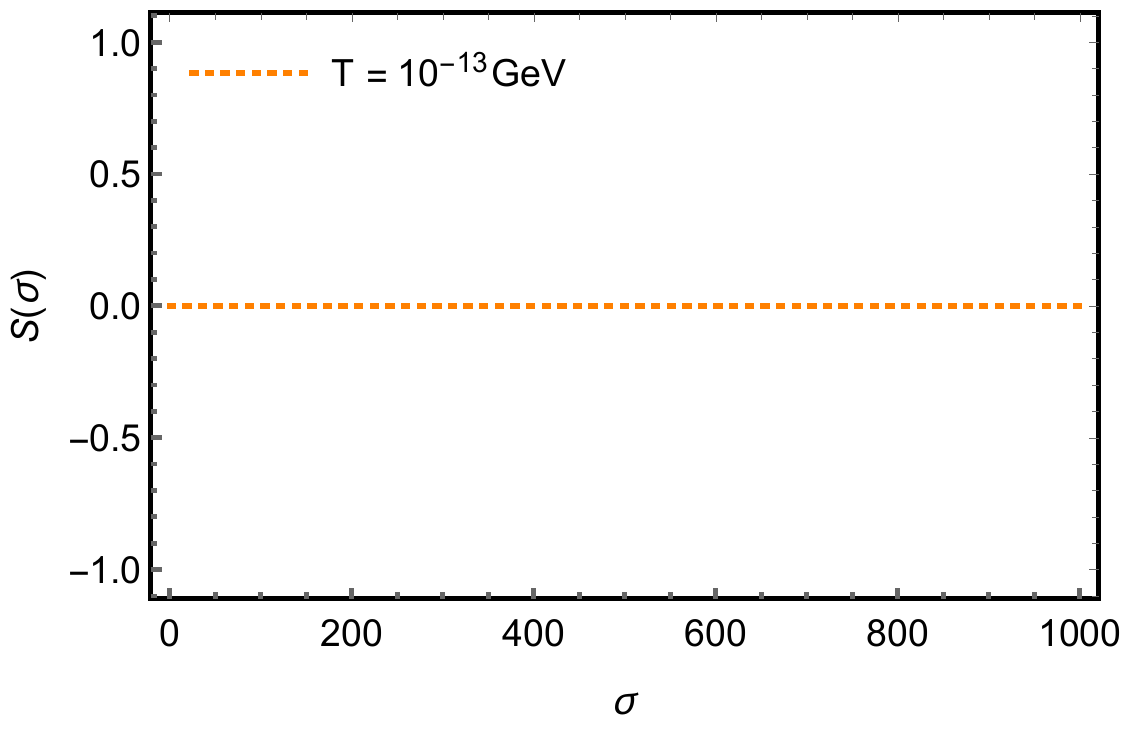}
\includegraphics[width=8cm,height=5cm]{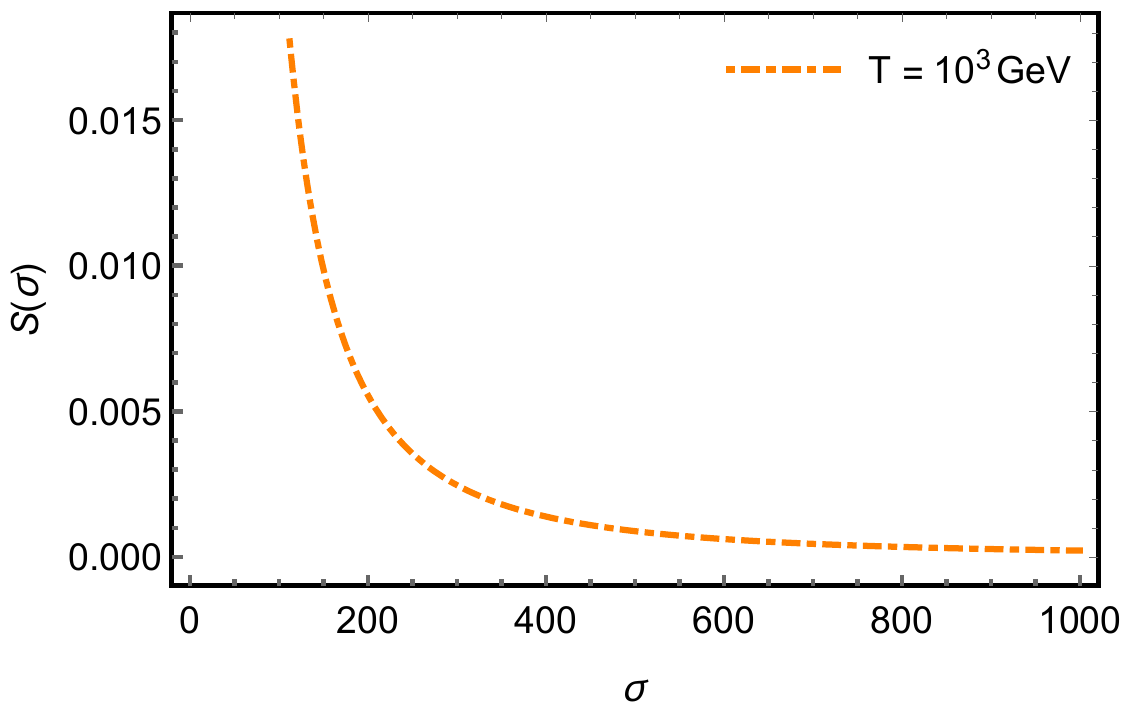}
\includegraphics[width=8cm,height=5cm]{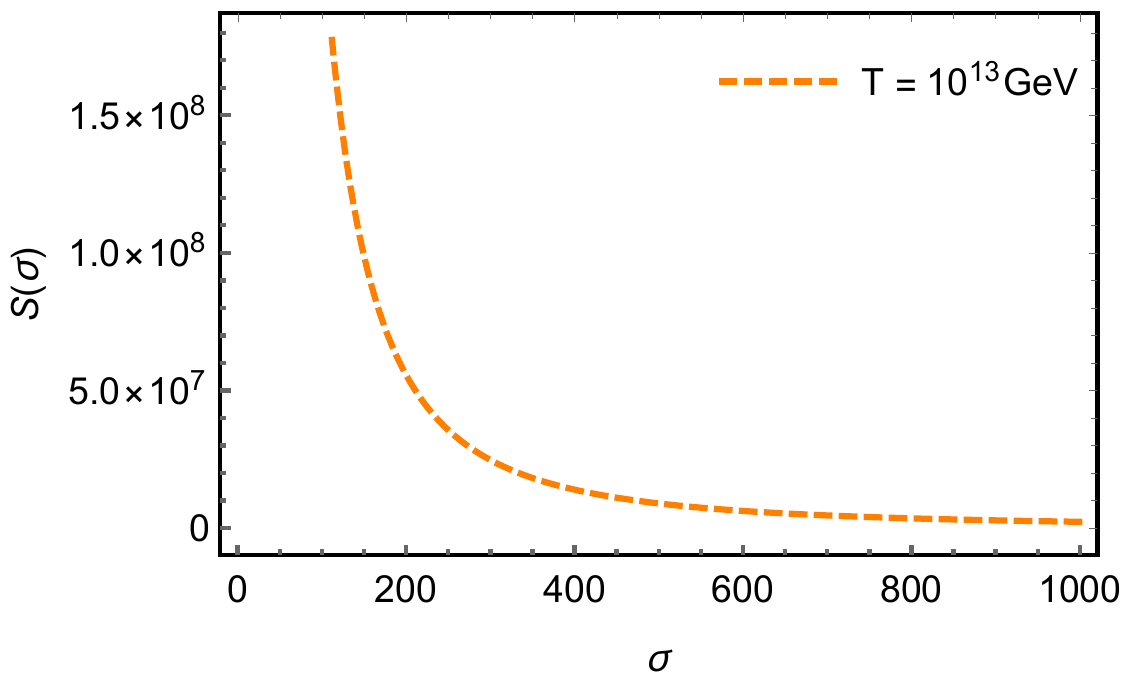}
\caption{The plots show the modification to the entropy $S(\sigma)$ as a function of $\sigma$ (its unit is GeV$^{-1}$) considering the temperatures of cosmic microwave background (top left), electroweak scenario (top right), and the early inflationary universe (bottom).}
\label{entropies1}
\end{figure}

\begin{figure}[ht]
\centering
\includegraphics[width=8cm,height=5cm]{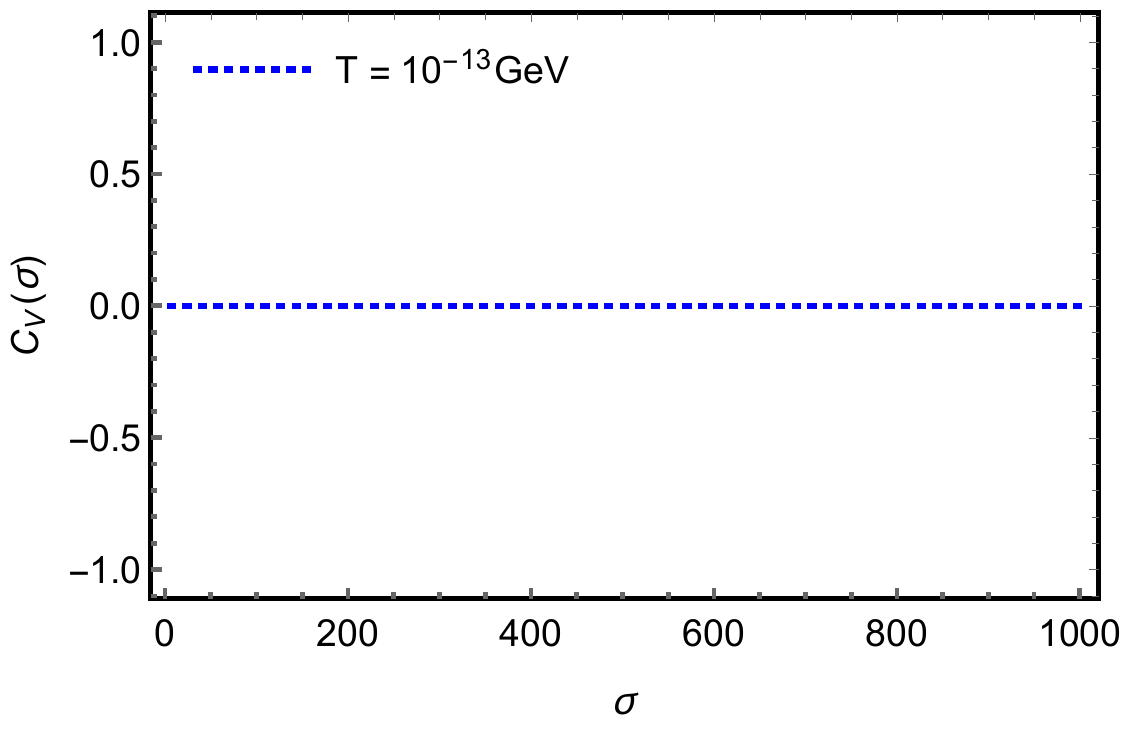}
\includegraphics[width=8cm,height=5cm]{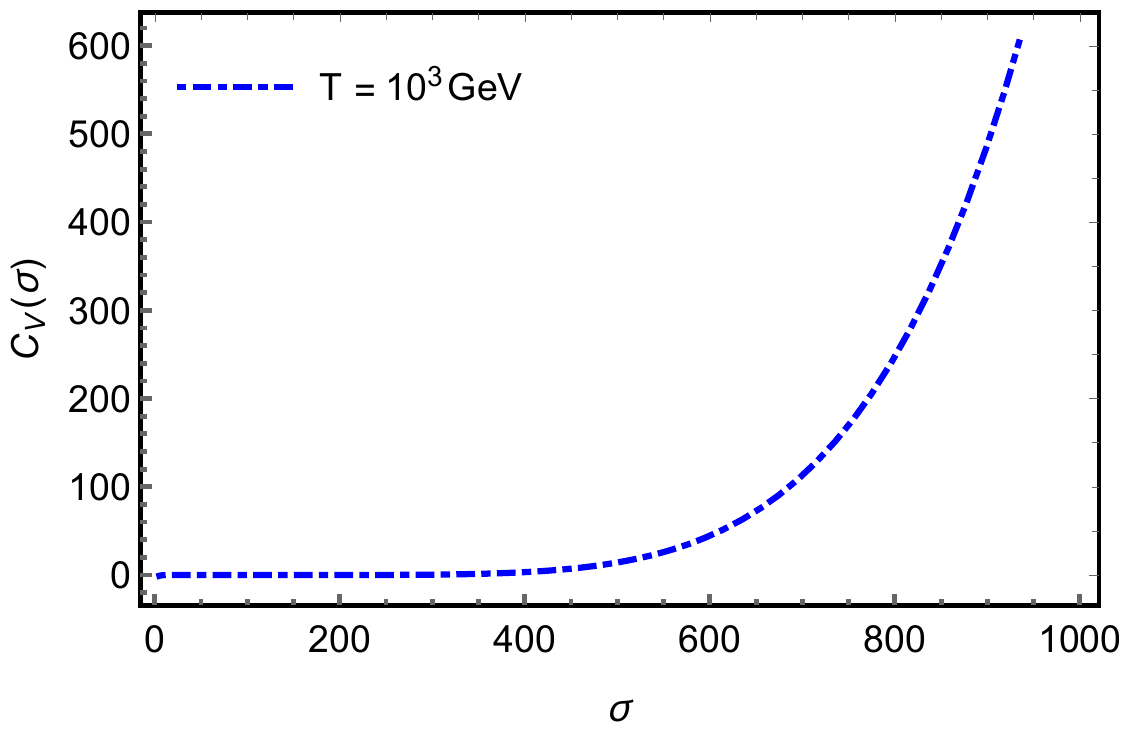}
\includegraphics[width=8cm,height=5cm]{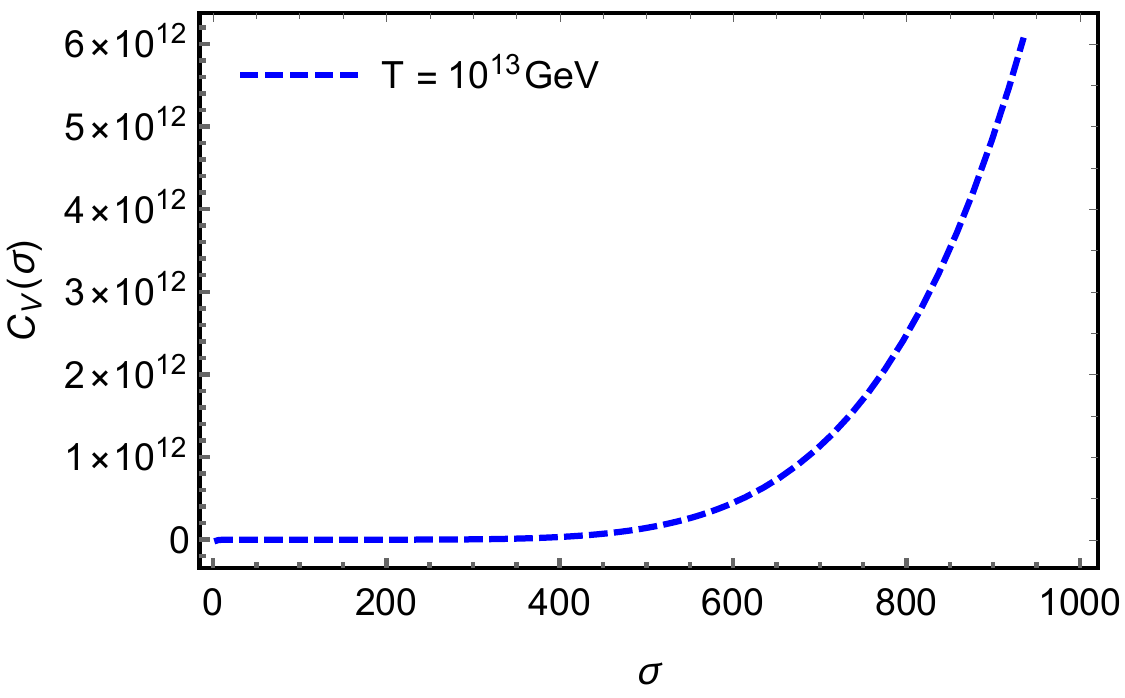}
\caption{The plots show the modification to the heat capacity $C_{V}(\sigma)$ as a function of $\sigma$ (its unit is GeV$^{-1}$) considering the temperatures of cosmic microwave background (top left), electroweak scenario (top right), and the early inflationary universe (bottom).}
\label{heatcapacities1}
\end{figure}

Using the expressions above, we can first obtain the 
Helmholtz free energy as being
\begin{eqnarray}
F(\beta,\sigma) &= & \frac{V}{\beta\pi^{2}}  \int^{\infty}_{0}
 \left(\frac{\sqrt[3]{\frac{2}{3}} \left(\frac{54 \sqrt{3} \sigma^{16} E^3}{\sqrt{4 \sigma ^{12}+27 \sigma^{16} E^4}} + 18 \sigma^8 E\right)}{3 \left(\sqrt{3} \sqrt{4 \sigma ^{12}+27 \sigma^{16} E^4}+9 \sigma^8 E^2\right)^{4/3}} \right.\\
&+&\left. 
\frac{\frac{54 \sqrt{3} \sigma^{16} E^3}{\sqrt{4 \sigma^{12}+27 \sigma^{16} E^4}}+18 \sigma^8 E}{3 \sqrt[3]{2} 3^{2/3} \sigma^4 \left(\sqrt{3} \sqrt{4 \sigma^{12}+27 \sigma^{16} E^4}+9 \sigma^8 E^2\right)^{2/3}}\right) \times \mathrm{ln} \left(  1- e^{-\beta E} \right) \nonumber\\
&\times& \sqrt{\frac{\sqrt[3]{\sqrt{3} \sqrt{4 \sigma ^{12}+27 \sigma^{16} E^4}+9 \sigma^8 E^2}}{\sqrt[3]{2} 3^{2/3} \sigma^4}-\frac{\sqrt[3]{\frac{2}{3}}}{\sqrt[3]{\sqrt{3} \sqrt{4 \sigma^{12}+27 \sigma^{16} E^4}+9 \sigma^8 E^2}}}\,\mathrm{d}E, \label{hlmontz1}\nonumber
\end{eqnarray}
the entropy
\begin{eqnarray}
S(\beta,\sigma) &= & -\frac{V k_{B}}{\pi^{2}}  \int^{\infty}_{0}
 \left(\frac{\sqrt[3]{\frac{2}{3}} \left(\frac{54 \sqrt{3} \sigma^{16} E^3}{\sqrt{4 \sigma ^{12}+27 \sigma^{16} E^4}} + 18 \sigma^8 E\right)}{3 \left(\sqrt{3} \sqrt{4 \sigma ^{12}+27 \sigma^{16} E^4}+9 \sigma^8 E^2\right)^{4/3}} \right.\\
&+&\left. 
\frac{\frac{54 \sqrt{3} \sigma^{16} E^3}{\sqrt{4 \sigma^{12}+27 \sigma^{16} E^4}}+18 \sigma^8 E}{3 \sqrt[3]{2} 3^{2/3} \sigma^4 \left(\sqrt{3} \sqrt{4 \sigma^{12}+27 \sigma^{16} E^4}+9 \sigma^8 E^2\right)^{2/3}}\right) \times \mathrm{ln} \left(  1- e^{-\beta E} \right) \nonumber\\
& \times& \sqrt{\frac{\sqrt[3]{\sqrt{3} \sqrt{4 \sigma ^{12}+27 \sigma^{16} E^4}+9 \sigma^8 E^2}}{\sqrt[3]{2} 3^{2/3} \sigma^4}-\frac{\sqrt[3]{\frac{2}{3}}}{\sqrt[3]{\sqrt{3} \sqrt{4 \sigma^{12}+27 \sigma^{16} E^4}+9 \sigma^8 E^2}}} \nonumber\\
& +& \frac{V\beta k_{B}}{\pi^{2}}  \int^{\infty}_{0}
E \left(\frac{\sqrt[3]{\frac{2}{3}} \left(\frac{54 \sqrt{3} \sigma^{16} E^3}{\sqrt{4 \sigma ^{12}+27 \sigma^{16} E^4}} + 18 \sigma^8 E\right)}{3 \left(\sqrt{3} \sqrt{4 \sigma ^{12}+27 \sigma^{16} E^4}+9 \sigma^8 E^2\right)^{4/3}} \right. \nonumber\\
&+&\left. 
\frac{\frac{54 \sqrt{3} \sigma^{16} E^3}{\sqrt{4 \sigma^{12}+27 \sigma^{16} E^4}}+18 \sigma^8 E}{3 \sqrt[3]{2} 3^{2/3} \sigma^4 \left(\sqrt{3} \sqrt{4 \sigma^{12}+27 \sigma^{16} E^4}+9 \sigma^8 E^2\right)^{2/3}}\right) \times \frac{e^{-\beta E}}{1 -e^{-\beta E}} \nonumber\\
& \times &\sqrt{\frac{\sqrt[3]{\sqrt{3} \sqrt{4 \sigma ^{12}+27 \sigma^{16} E^4}+9 \sigma^8 E^2}}{\sqrt[3]{2} 3^{2/3} \sigma^4}-\frac{\sqrt[3]{\frac{2}{3}}}{\sqrt[3]{\sqrt{3} \sqrt{4 \sigma^{12}+27 \sigma^{16} E^4}+9 \sigma^8 E^2}}} \,\mathrm{d}E, \nonumber \label{entropy1}
\end{eqnarray}
and, lastly, the heat capacity
\begin{eqnarray}
C_{V}(\beta,\sigma) &= & \frac{Vk_{B}\beta^{2}}{\pi^{2}}   \int^{\infty}_{0}
 E^{2} \left(\frac{\sqrt[3]{\frac{2}{3}} \left(\frac{54 \sqrt{3} \sigma^{16} E^3}{\sqrt{4 \sigma ^{12}+27 \sigma^{16} E^4}} + 18 \sigma^8 E\right)}{3 \left(\sqrt{3} \sqrt{4 \sigma ^{12}+27 \sigma^{16} E^4}+9 \sigma^8 E^2\right)^{4/3}} \right.\\
&+&\left. 
\frac{\frac{54 \sqrt{3} \sigma^{16} E^3}{\sqrt{4 \sigma^{12}+27 \sigma^{16} E^4}}+18 \sigma^8 E}{3 \sqrt[3]{2} 3^{2/3} \sigma^4 \left(\sqrt{3} \sqrt{4 \sigma^{12}+27 \sigma^{16} E^4}+9 \sigma^8 E^2\right)^{2/3}}\right) \times \frac{e^{-2\beta E}}{(1 -e^{-\beta E})^{2}} \nonumber\\
& \times& \sqrt{\frac{\sqrt[3]{\sqrt{3} \sqrt{4 \sigma ^{12}+27 \sigma^{16} E^4}+9 \sigma^8 E^2}}{\sqrt[3]{2} 3^{2/3} \sigma^4}-\frac{\sqrt[3]{\frac{2}{3}}}{\sqrt[3]{\sqrt{3} \sqrt{4 \sigma^{12}+27 \sigma^{16} E^4}+9 \sigma^8 E^2}}} \nonumber\\
&+& \frac{ V k_{B}\beta^{2}}{\pi^{2}}  \int^{\infty}_{0}
E^{2} \left(\frac{\sqrt[3]{\frac{2}{3}} \left(\frac{54 \sqrt{3} \sigma^{16} E^3}{\sqrt{4 \sigma ^{12}+27 \sigma^{16} E^4}} + 18 \sigma^8 E\right)}{3 \left(\sqrt{3} \sqrt{4 \sigma ^{12}+27 \sigma^{16} E^4}+9 \sigma^8 E^2\right)^{4/3}} \right.\nonumber\\
&+&\left. 
\frac{\frac{54 \sqrt{3} \sigma^{16} E^3}{\sqrt{4 \sigma^{12}+27 \sigma^{16} E^4}}+18 \sigma^8 E}{3 \sqrt[3]{2} 3^{2/3} \sigma^4 \left(\sqrt{3} \sqrt{4 \sigma^{12}+27 \sigma^{16} E^4}+9 \sigma^8 E^2\right)^{2/3}}\right) \times \frac{e^{-\beta E}}{1 -e^{-\beta E}} \nonumber\\
& \times& \sqrt{\frac{\sqrt[3]{\sqrt{3} \sqrt{4 \sigma ^{12}+27 \sigma^{16} E^4}+9 \sigma^8 E^2}}{\sqrt[3]{2} 3^{2/3} \sigma^4}-\frac{\sqrt[3]{\frac{2}{3}}}{\sqrt[3]{\sqrt{3} \sqrt{4 \sigma^{12}+27 \sigma^{16} E^4}+9 \sigma^8 E^2}}} \,\mathrm{d}E. \nonumber\label{entropy1a}
\end{eqnarray}
Their behaviors are shown in Figs. \ref{fs1}, \ref{entropies1}, and \ref{heatcapacities1} respectively. Moreover, the thermal quantities were also calculated \cite{oliveira2019thermodynamic,oliveira2020thermodynamic,reis2020does,lobo2020effects,lobo2020thermal,amelino2017thermal} into different contexts. In the context of CMB, all of them turned out to have no contribution to our calculations. For Helmholtz free energy, we obtained decreasing curves with an expressive curvature when $\sigma$ increases for both electroweak and inflationary cases. The entropy, on the other hand, showed a decreasing behavior for different values of $\sigma$ for both electroweak and inflationary cases. It is worth mentioning that such behavior does not imply an instability since the entropy is still an increasing function when it is analyzed against the temperature for fixed values of $\sigma$ -- as it should be. Lastly, the heat capacity exhibited a strong increasing behavior when $\sigma$ started to change for both cases as well, i.e., the CMB and the inflationary temperatures. In principle, this fact might signalize the possibility of some phase transition at some large $\sigma$; nevertheless, this hypothesis requires further investigation.

Whenever we are dealing with the thermodynamic systems, one question naturally arises: what is the form of the equation of state when the parameter $\sigma$, which characterizes the magnitude of Lorentz symmetry breaking, is taken into account? To answer such a question, we must start with the following relation:
\ie
\mathrm{d}F = - S \,\mathrm{d}T - p\,\mathrm{d}V,
\fe
which, immediately, implies 

\begin{eqnarray}
p= - \left(\frac{\partial F}{\partial V} \right)_{T} &= &- \frac{1}{\beta\pi^{2}}  \int^{\infty}_{0}
 \left(\frac{\sqrt[3]{\frac{2}{3}} \left(\frac{54 \sqrt{3} \sigma^{16} E^3}{\sqrt{4 \sigma ^{12}+27 \sigma^{16} E^4}} + 18 \sigma^8 E\right)}{3 \left(\sqrt{3} \sqrt{4 \sigma ^{12}+27 \sigma^{16} E^4}+9 \sigma^8 E^2\right)^{4/3}} \right.\\
&+&\left. 
\frac{\frac{54 \sqrt{3} \sigma^{16} E^3}{\sqrt{4 \sigma^{12}+27 \sigma^{16} E^4}}+18 \sigma^8 E}{3 \sqrt[3]{2} 3^{2/3} \sigma^4 \left(\sqrt{3} \sqrt{4 \sigma^{12}+27 \sigma^{16} E^4}+9 \sigma^8 E^2\right)^{2/3}}\right) \times \mathrm{ln} \left(  1- e^{-\beta E} \right) \nonumber\\
&\times& \sqrt{\frac{\sqrt[3]{\sqrt{3} \sqrt{4 \sigma ^{12}+27 \sigma^{16} E^4}+9 \sigma^8 E^2}}{\sqrt[3]{2} 3^{2/3} \sigma^4}-\frac{\sqrt[3]{\frac{2}{3}}}{\sqrt[3]{\sqrt{3} \sqrt{4 \sigma^{12}+27 \sigma^{16} E^4}+9 \sigma^8 E^2}}}\,\mathrm{d}E. \nonumber
\label{pressure}
\end{eqnarray}

\begin{figure}[ht]
\centering
\includegraphics[width=9cm,height=6cm]{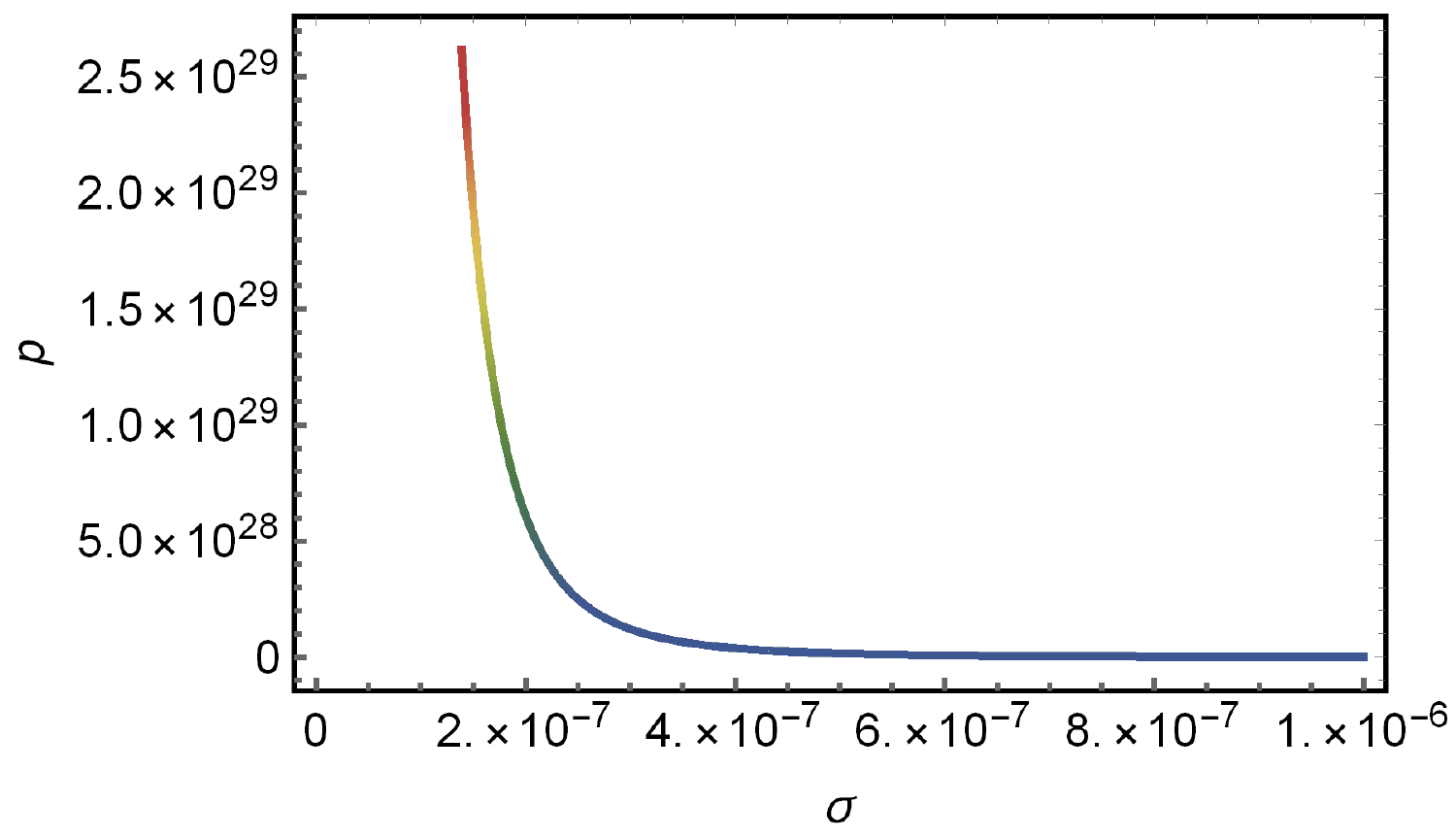}
\caption{This figure shows the behavior of the equation of states when the high temperature limit, namely $\frac{\beta}{\sigma} \ll 1$, is taken into account.}
\label{highlimit}
\end{figure}

\begin{figure}[ht]
\centering
\includegraphics[width=9cm,height=6cm]{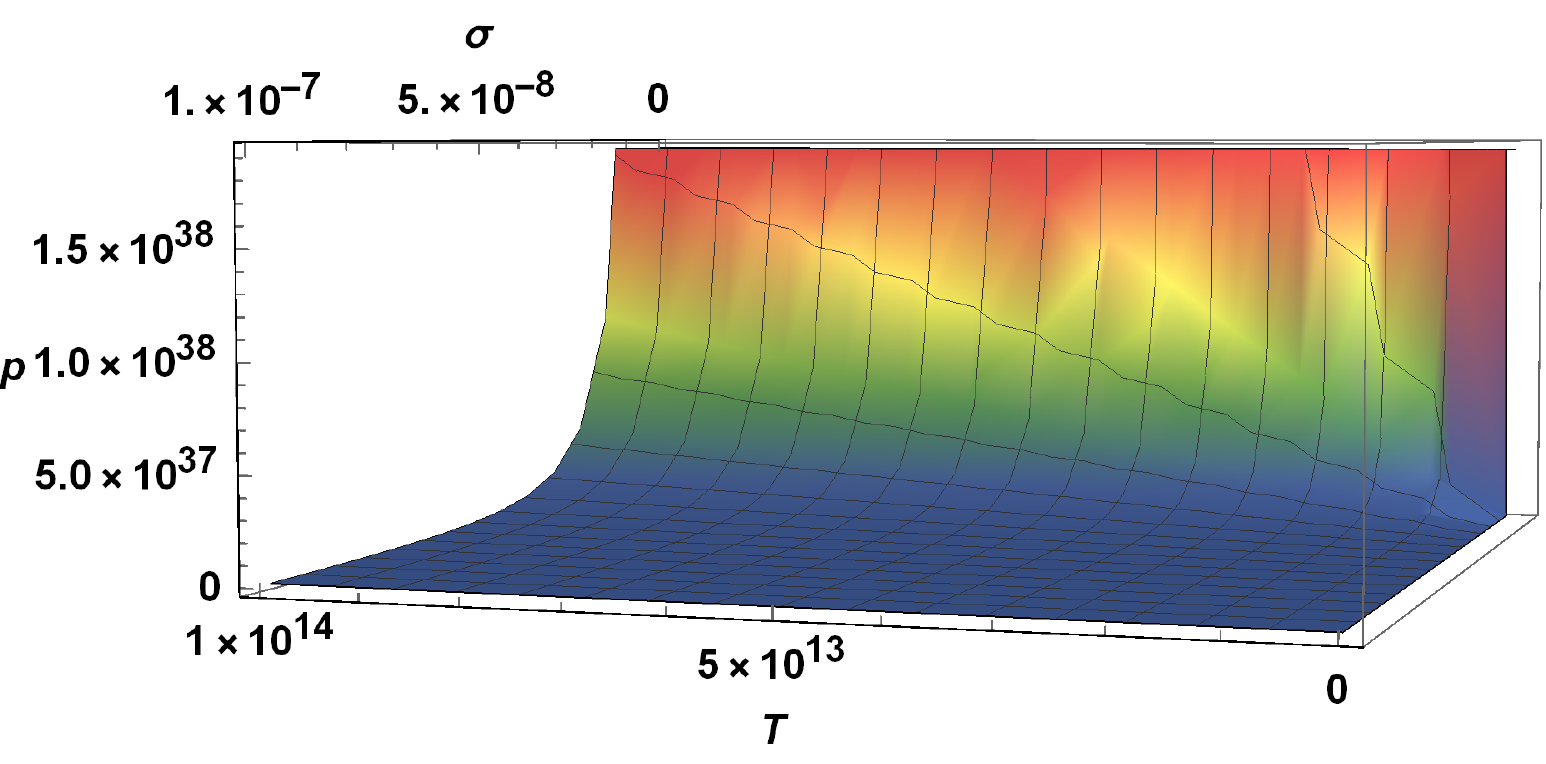}
\caption{This figure shows the behavior of the equation of states when the low temperature limit, namely $\frac{\beta}{\sigma} \gg 1$, is taken into account.}
\label{lowlimit}
\end{figure}

Here, we focus on the study of the behavior of pressure -- which leads to the equation of states. After introducing new dimensionless variable $t=\sigma E$, we get
	\begin{eqnarray}
		p&=&\frac{1}{\sigma^3\beta\pi^2}\int_0^{\infty} \mathrm{d}t \,\mathrm{ln}(1-e^{-\frac{\beta}{\sigma}t})F(t);\nonumber\\
		F(t)&=&-2^{1/6}3^{-7/6}\frac{t}{\sqrt{4+27t^4}}\left[(\chi(t))^{-1/3}+\frac{1}{\sqrt[3]{18}}(\chi(t))^{1/3}\right]\cdot\sqrt{\frac{(\chi(t))^{2/3}-12^{1/3}}{(\chi(t))^{1/3}}};
		\nonumber\\
		\chi(t)&=&\sqrt{3}\cdot\sqrt{4+27t^4}+9t^2.	
	\end{eqnarray}
With these above expressions, let us examine some limits. Initially, at the very high temperature limit, namely $\frac{\beta}{\sigma}=\frac{1}{\sigma T}\ll 1$, we perform the expansion in the Taylor series. After that, only the first two terms are considered, which yields $p=\frac{1}{\sigma^3\beta\pi^2}\frac{\beta}{\sigma}\int_0^{\infty} \mathrm{d}t  F(t)t =\frac{c_h}{\sigma^4\pi^2}$, where $c_h=\int_0^{\infty} \mathrm{d}t  F(t)t$, i.e., at $T\to\infty$ ($\beta\to 0$), the pressure tends to a constant. Such behavior is displayed in Fig. \ref{highlimit}; next, at the natural limit of low Lorentz symmetry breaking or low temperature, one has $\frac{\beta}{\sigma}\gg 1$. In this case, the exponential is strongly suppressed, and  one has $p=\frac{c_l}{\sigma^3\beta\pi^2}$, where $c_l=\int_0^{\infty} \mathrm{d}t F(t)$, i.e., the pressure grows linearly with the temperature. Thereby, its corresponding behavior is shown in Fig. \ref{lowlimit}.

\section{Thermodynamical aspects of CPT-odd higher-derivative LV theory}

In this section, let us consider the theory described by the following dispersion relation \cite{rosati2015planck}:
\ie
E^{2} +\alpha lE^{3} +\beta lE{\bf{k}}^{2}={\bf k}^2+m^2. \label{disper}
\fe
where $l$ is a parameter characterizing the intensity of the Lorentz symmetry breaking. It is clear that in the limit $l \rightarrow 0$, the standard massive dispersion relation $ E^{2}={\bf{k}}^{2} + m^{2}$ is recovered. Such a relation, being similar to relations studied in Ref. \cite{amelino2001testable}, can arise e.g. in a scalar field theory with the higher-derivative quadratic Lagrangian of the scalar field looking like ${\cal L}=\frac{1}{2}\phi(\Box+m^2+\rho^{\mu\nu\lambda}\partial_{\mu}\partial_{\nu}\partial_{\rho})\phi$, with $\rho^{\mu\nu\lambda}$ being a completely symmetric third-rank tensor whose only non-zero components are $\rho^{000}=\beta l$, and $\rho^{0ij}=\frac{1}{3}\alpha l\delta^{ij}$. In principle, it is natural to expect that such a relation, in the massless case, can also arise in a specific higher-derivative LV extension of QED. In particular, the dispersion relation displayed in Eq. (\ref{disper}) can be applied to understand how Planck-scale effects may affect translation transformations. This is relevant due to the fact that it carries the information on the distance between source and detector, and it factors in the interplay between quantum-spacetime effects and the curvature of spacetime \cite{amelino2016icecube,rosati2015planck}. In the literature, some propositions are addressed in the context of gamma-ray-burst neutrinos and photons \cite{amelino2017vacuo}, and IceCube and GRB neutrinos \cite{amelino2016icecube}.

\begin{figure}[ht]
\centering
\includegraphics[width=8cm,height=5cm]{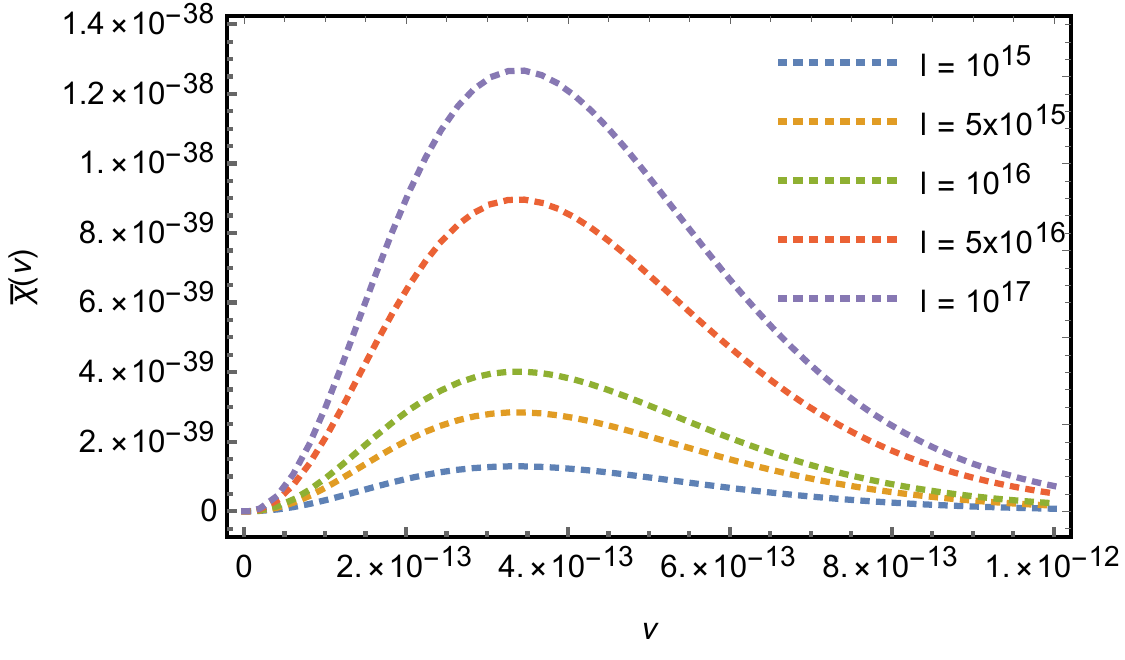}
\includegraphics[width=8cm,height=5cm]{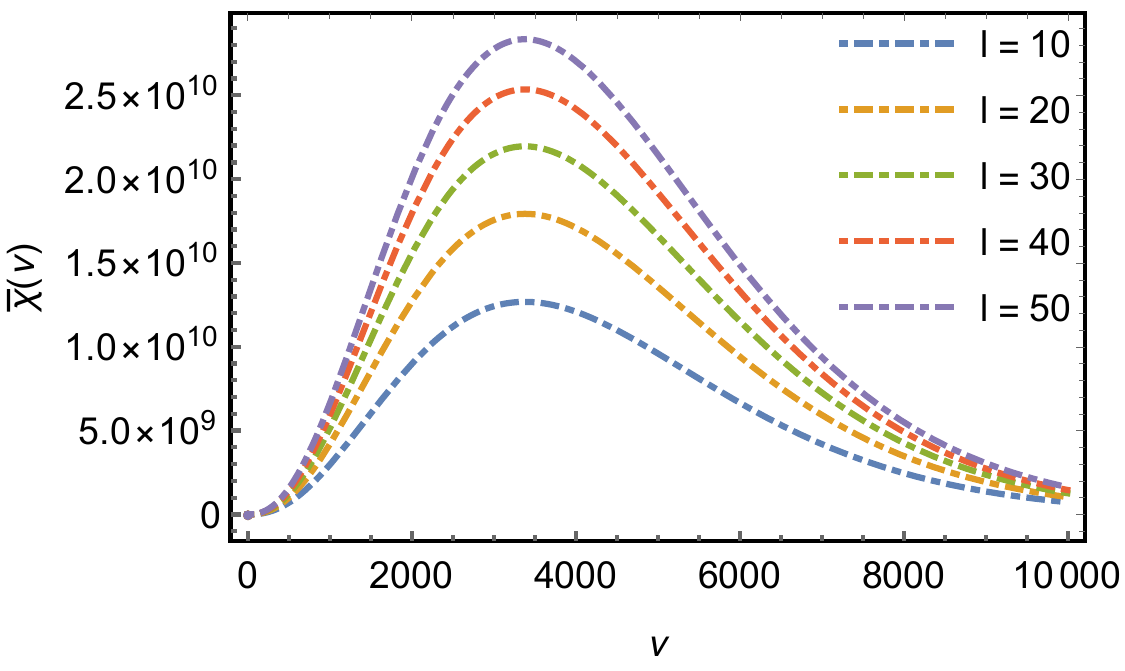}
\includegraphics[width=8cm,height=5cm]{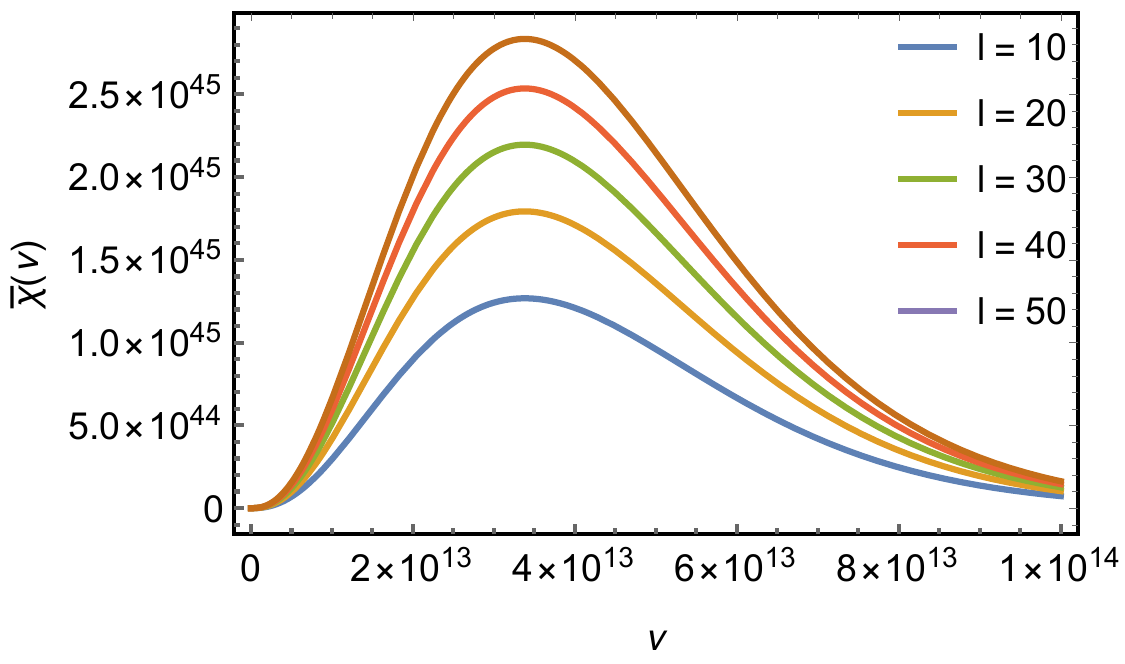}
\caption{The plots show how the spectral radiance $\Bar{\chi}(\nu)$ changes as a function of frequency $\nu$ and $l (\text{whose dimension is}\,\, m\cdot kg^{-1/2}\cdot s^{-1})$ for three different cases. The top left (dotted) is configuration to the cosmic microwave background, i.e., $\beta = 10^{13}$ GeV; the top right (dot-dashed) is ascribed to the electroweak configuration, i.e., $\beta = 10^{-3}$ GeV; the bottom plot shows the black body radiation to the inflationary period of the Universe, i.e., $\beta = 10^{-13}$ GeV. }
\label{spectral-radiances}
\end{figure}

Thereby, we restrict ourselves to a particular massless case, i.e., we set $m=0, \alpha=1$ and $\beta= 1$. In this sense, it is convenient to rewrite Eq. (\ref{disper}) as
\ie
{\bf{k}}^{2} = \frac{E^{2} + lE^{3} }{1-lE} \label{dr},
\fe
where, analogously with the previous section, the accessible states can be derived: 
\ie
\Bar{\Omega}(l) = \frac{\Gamma}{2\pi^{2}} \int^{\infty}_{0}   \left(  \frac{E^{2}+lE^{3}}{1 - lE} \right)^{1/2}  \left[ \frac{(2E + 3lE^{2})(1-lE) + l(E^{2}+lE^{3})}{(1-lE)^{2}}   \right]    \mathrm{d}E.
\fe
With this, we are able to write down the corresponding partition function as follows
\begin{eqnarray}
&&\mathrm{ln}\left[ \Bar{Z}(\beta,l)\right] =\\
&=& - \frac{\Gamma}{\pi^{2}} \int^{\infty}_{0}   \left(  \frac{E^{2}+lE^{3}}{1 - lE} \right)^{1/2}  \left[ \frac{(2E + 3lE^{2})(1-lE) + l(E^{2}+lE^{3})}{(1-lE)^{2}}   \right] \mathrm{ln} \left(  1- e^{-\beta E} \right)
\mathrm{d}E.\nonumber
\label{partition2}
\end{eqnarray}
Using Eq. (\ref{partition2}), we can obtain the thermodynamic functions per volume $\Gamma$ as well. Here, we provide the calculation of Helmholtz free energy $\Bar{F}(\beta,l)$, mean energy $\Bar{U}(\beta,l)$, entropy $\Bar{S}(\beta,l)$, and heat capacity $\Bar{C}_{V}(\beta,l)$.
Let us start with the mean energy
\ie
\Bar{U}(\beta,l) = \frac{1}{\pi^{2}}  \int^{\infty}_{0} 
E\left(  \frac{E^{2}+lE^{3}}{1 - lE} \right)^{1/2}  \left[ \frac{(2E + 3lE^{2})(1-lE) + l(E^{2}+lE^{3})}{(1-lE)^{2}}   \right]
\frac{e^{-\beta E}}{\left(  1- e^{-\beta E} \right)} \mathrm{d}E, \label{meanenergy}
\fe
which implies the spectral radiance given by:
\begin{eqnarray}
\Bar{\chi}(l,\nu) &=& (h\nu)\left(  \frac{(h\nu)^{2}+l(h\nu)^{3}}{1 - l(h\nu)} \right)^{1/2}  \left[ \frac{(2(h\nu) + 3l(h\nu)^{2})(1-l(h\nu)) + l((h\nu)^{2}+l(h\nu)^{3})}{(1-l(h\nu))^{2}}   \right]
\times\nonumber\\&\times&
\frac{e^{-\beta (h\nu)}}{\left(  1- e^{-\beta (h\nu)} \right)}.
\label{spectralradiance}
\end{eqnarray}
The respective plots of these thermal quantities are presented in Fig. \ref{spectral-radiances}. Here, we show the black body radiation spectra for different values of $l$ corresponding to the Cosmic Microwave Background, electroweak epoch and inflationary era of the Universe. Moreover, the black body radiation shape is maintained for the three of them, differently what happened in our first example in the previous section. Note that when $l \rightarrow 0$, we recover the usual radiation constant of the \textit{Stefan–Boltzmann} law, namely, $u_{SB}= \alpha T^{4}$. In other words, we have
\ie
\alpha = \frac{1}{\pi^{2}}  \int^{\infty}_{0}  \frac{E^{3}\,e^{-\beta E}} {\left(  1- e^{-\beta E} \right)} \mathrm{d}E = \frac{\pi^{2}}{15}.
\label{radiance}
\fe

Furthermore, for the sake of examining how the parameter $l$ affects the correction to the \textit{Stefan–Boltzmann} law, we also consider
\ie
\Bar{\alpha} \equiv U(\beta,l) \beta^{4}. \label{sbl}
\fe
The plots are exhibited in Fig. \ref{alphas2} taking differently into account three scenarios, i.e., the temperatures of: CMB, electroweak epoch and the early inflationary era of the universe. Furthermore, the high-energy limit $2E+3lE^{2} \gg 1 - l E$ is also regarded. Here, when the CMB temperature is considered, we see a constant behavior of the curve. On the other hand, to the electroweak scenario, we obtain a monotonically increasing function as $l$ changes. Finally, in the inflationary era, we also have a stable model showing a rising behavior when $l$ increases.

In the same manner, the remaining thermodynamic functions can be explicitly computed:

\begin{figure}[ht]
\centering
\includegraphics[width=8cm,height=5cm]{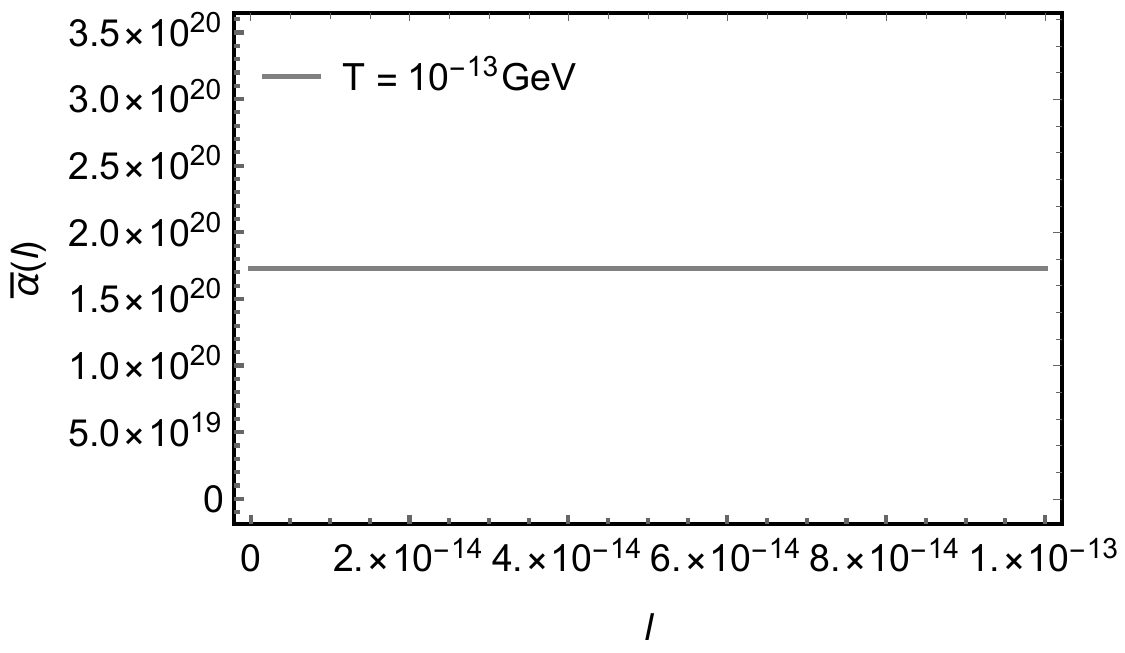}
\includegraphics[width=8cm,height=5cm]{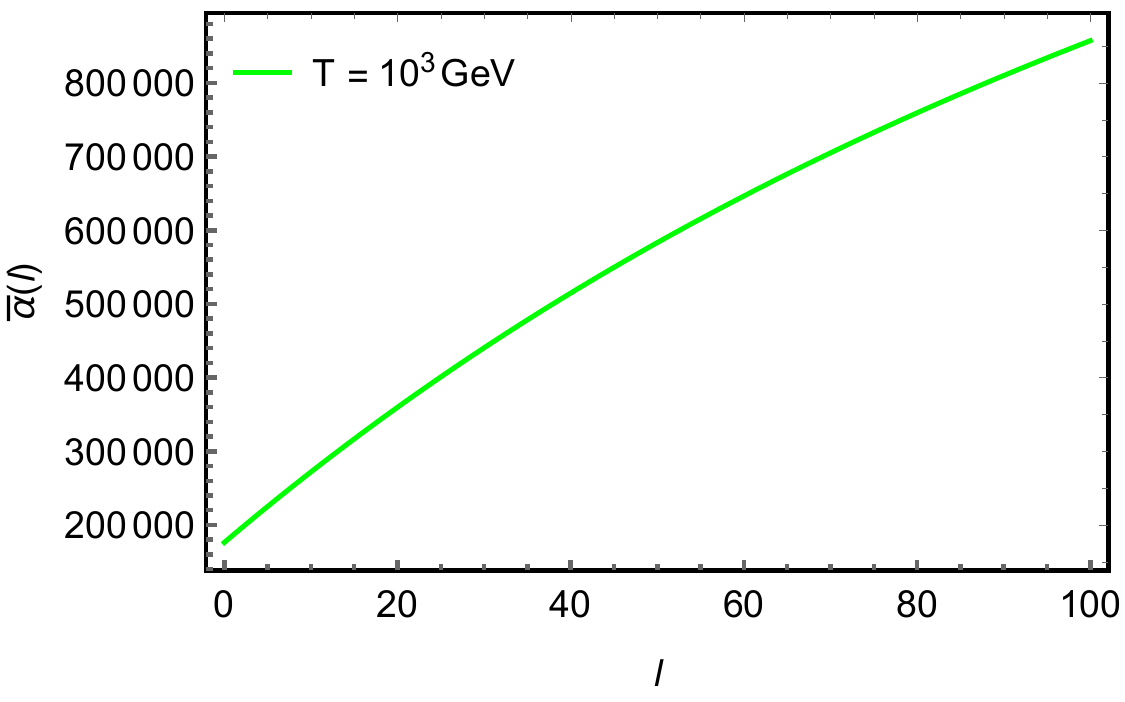}
\includegraphics[width=8cm,height=5cm]{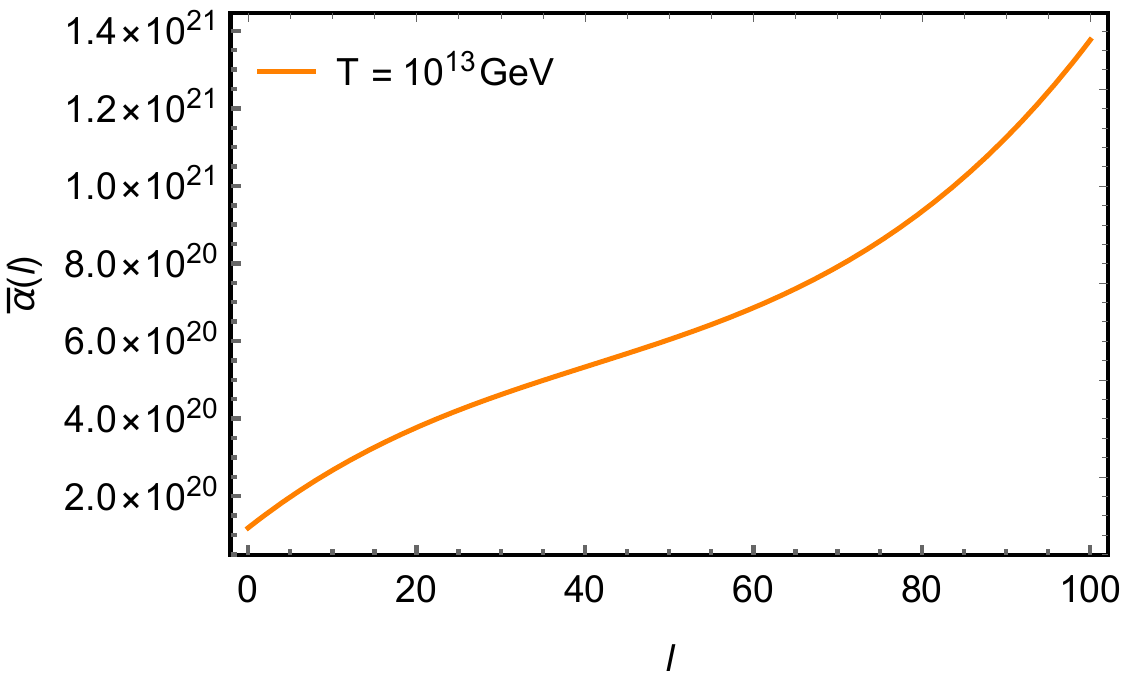}
\caption{The figure shows the correction to the \textit{Stefan–Boltzmann} law represented by parameter $\Bar{\alpha}$ as a function of $l (\text{whose dimension is}\,\, m\cdot kg^{-1/2}\cdot s^{-1})$ considering the temperatures of cosmic microwave background (top left), electroweak scenario (top right), and the early inflationary universe (bottom).}
\label{alphas2}
\end{figure}

\begin{figure}[ht]
\centering
\includegraphics[width=8cm,height=5cm]{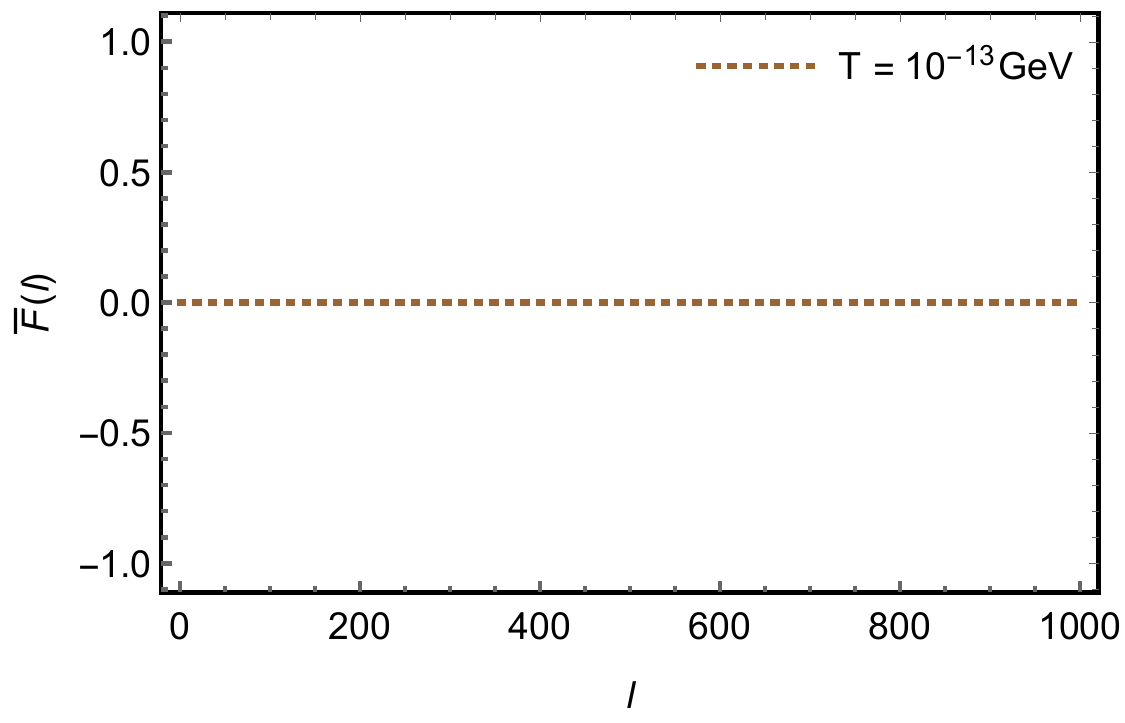}
\includegraphics[width=8cm,height=5cm]{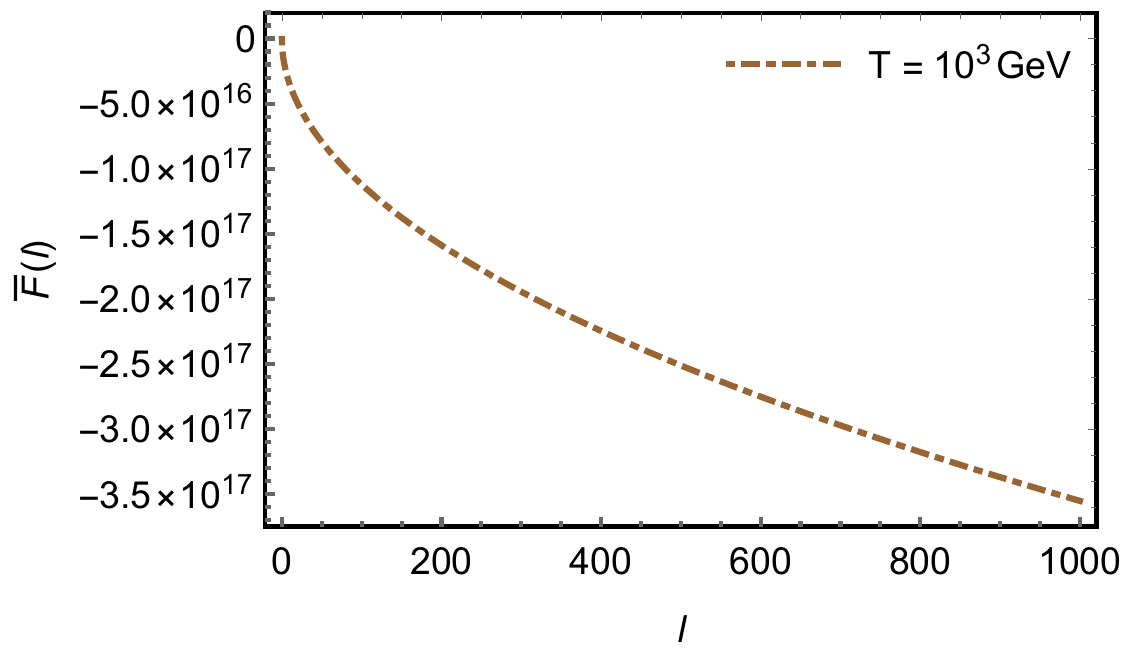}
\includegraphics[width=8cm,height=5cm]{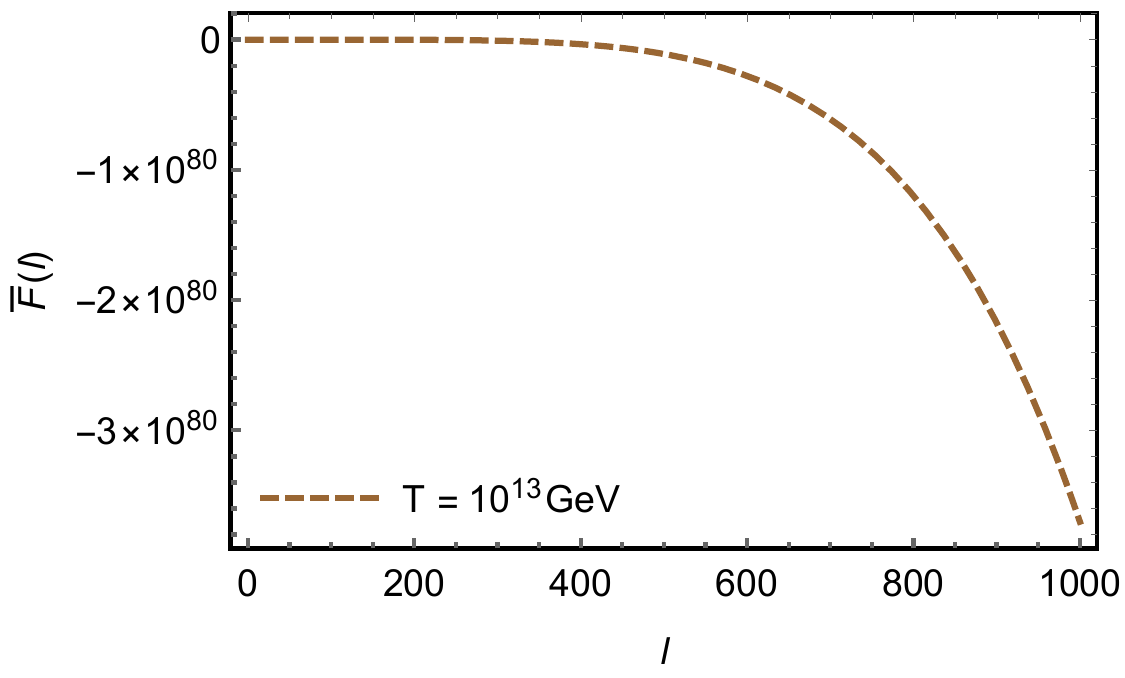}
\caption{The figure shows the modification of the Helmholtz free energy $\Bar{F}$ as a function of $l (\text{whose dimension is}\,\, m\cdot kg^{-1/2}\cdot s^{-1})$ considering the temperatures of cosmic microwave background (top left), electroweak scenario (top right), and the early inflationary universe (bottom).}
\label{fs2}
\end{figure}

\begin{figure}[ht]
\centering
\includegraphics[width=8cm,height=5cm]{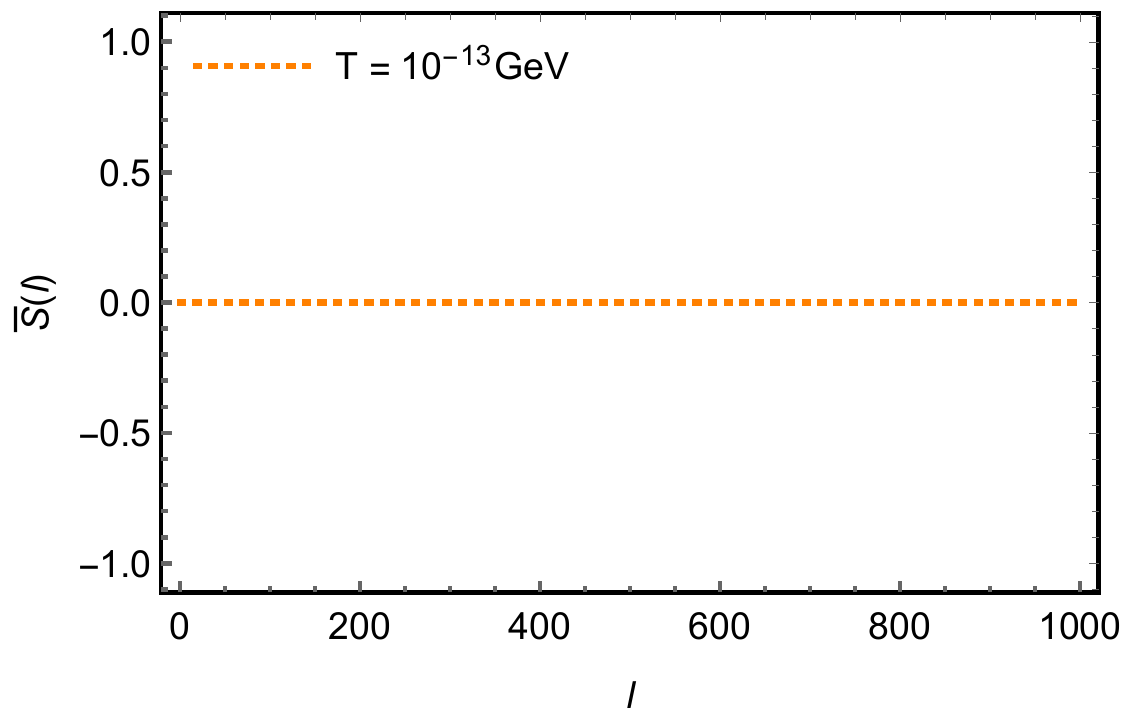}
\includegraphics[width=8cm,height=5cm]{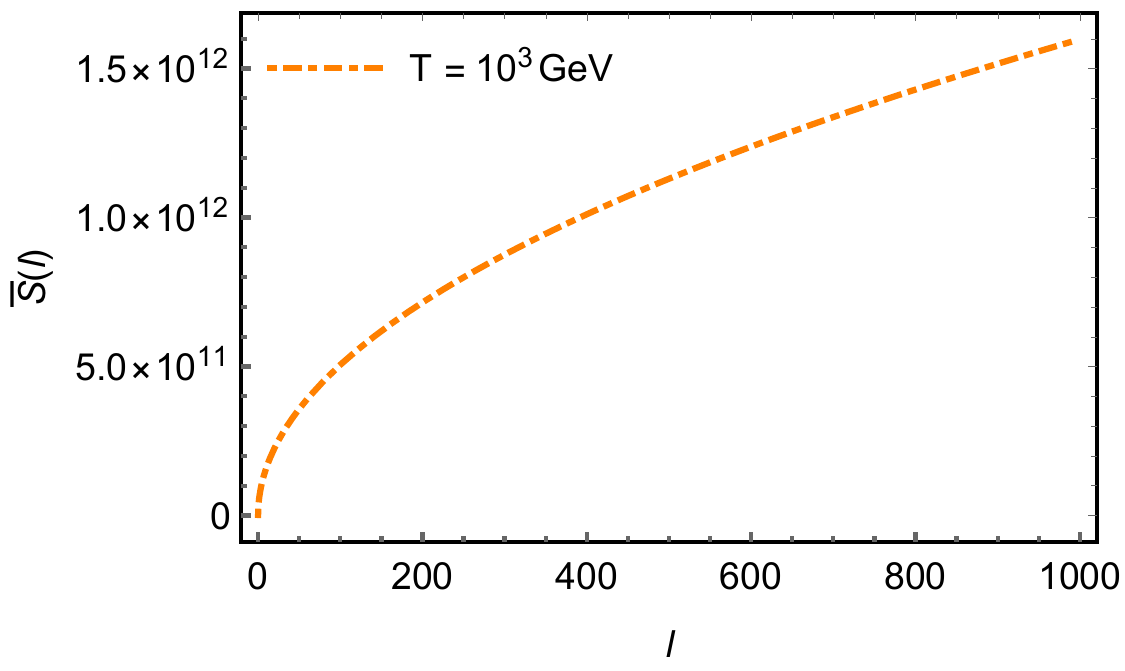}
\includegraphics[width=8cm,height=5cm]{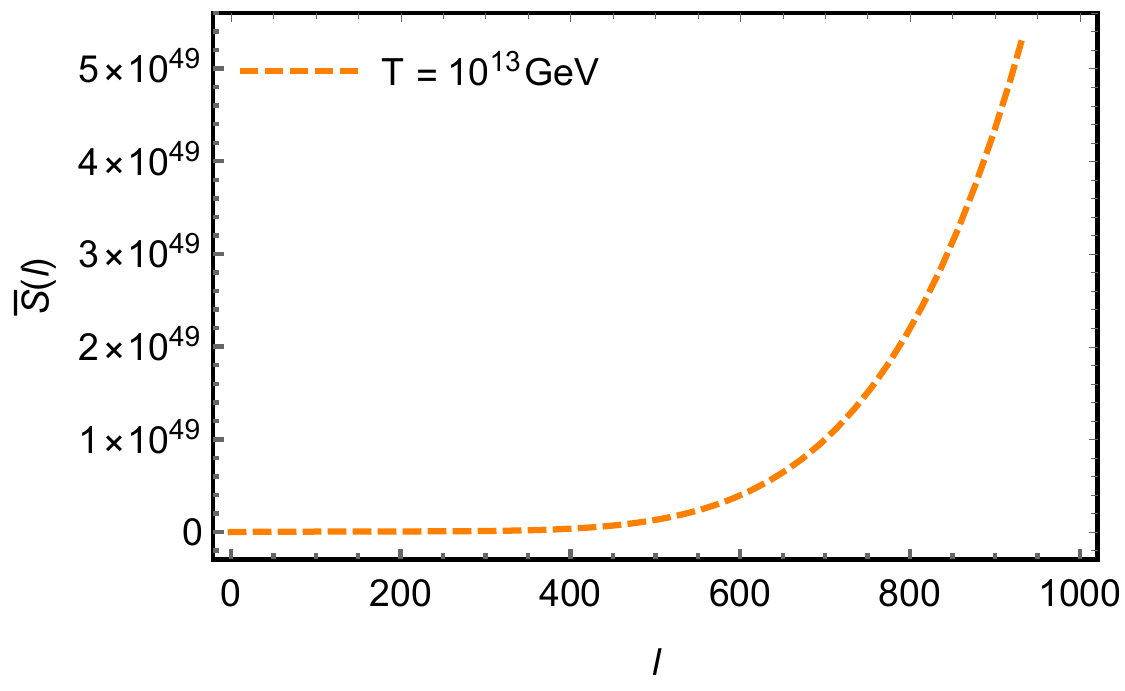}
\caption{The figure shows the modification of the entropy $\Bar{S}$ as a function of $l (\text{whose dimension is}\,\, m\cdot kg^{-1/2}\cdot s^{-1})$ considering the temperatures of cosmic microwave background (top left), electroweak scenario (top right) and the early inflationary universe (bottom).}
\label{entropies2}
\end{figure}

\begin{figure}[ht]
\centering
\includegraphics[width=8cm,height=5cm]{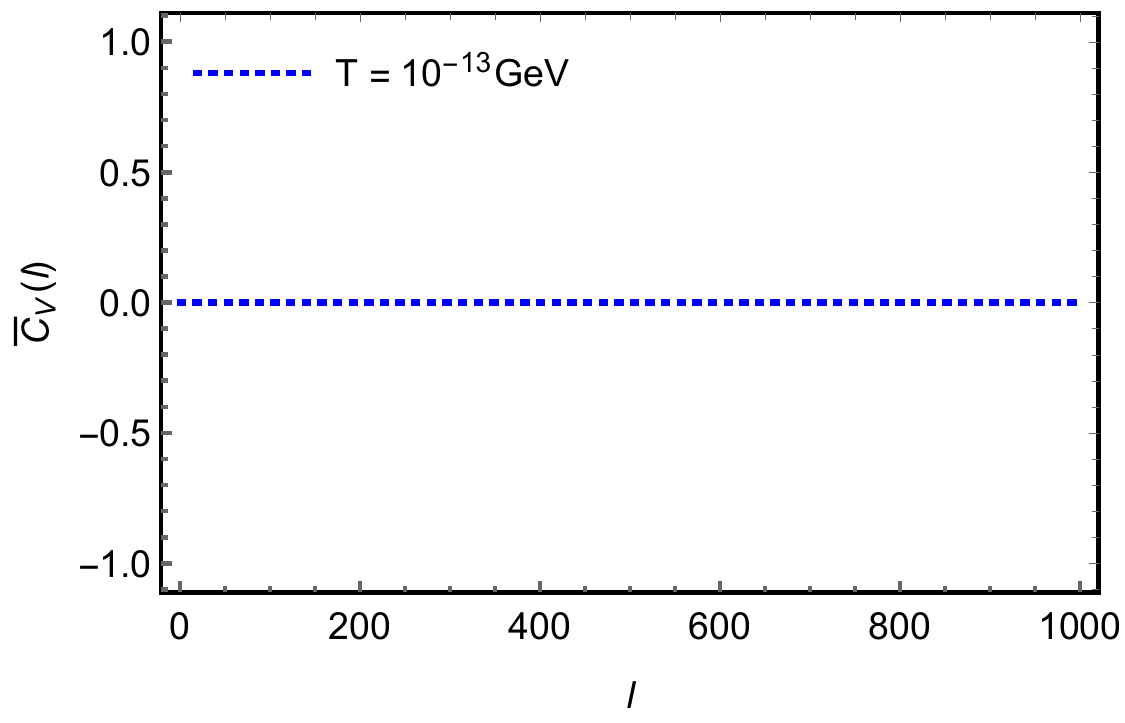}
\includegraphics[width=8cm,height=5cm]{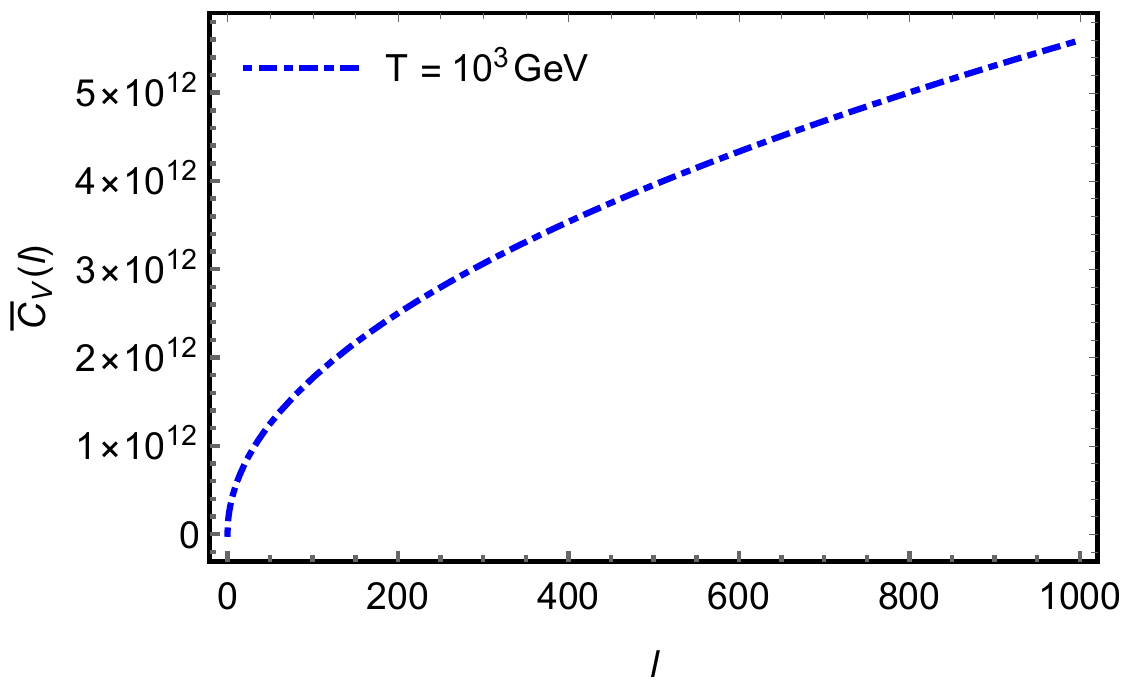}
\includegraphics[width=8cm,height=5cm]{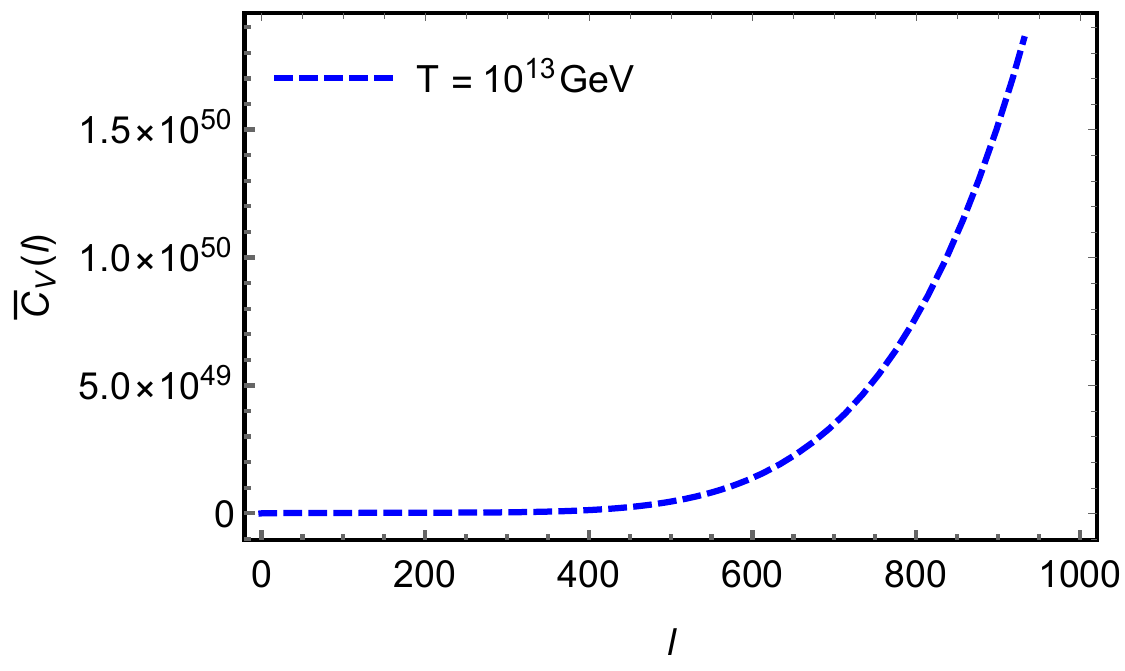}
\caption{The figure shows the modification of the heat capacity $\Bar{C}_{V}$ as a function of $l (\text{whose dimension is}\,\, m\cdot kg^{-1/2}\cdot s^{-1})$ considering the temperatures of cosmic microwave background (top left), electroweak scenario (top right) and the early inflationary universe (bottom).}
\label{heatcapacities2}
\end{figure}

\ie
\Bar{F}(\beta,l) = \frac{1}{ \pi^{2} \beta} \int^{\infty}_{0} \left(  \frac{E^{2}+lE^{3}}{1 - lE} \right)^{1/2}  \left[ \frac{(2E + 3lE^{2})(1-lE) + l(E^{2}+lE^{3})}{(1-lE)^{2}}   \right]\,\mathrm{ln}\left( 1-e^{-\beta E}\right) \mathrm{d}E, \label{helmontz2}
\fe
\ie
\begin{split}
\Bar{S}(\beta,l) = \frac{k_{B}}{ \pi^{2}} &\int_0^{\infty}\left\{ - \left(  \frac{E^{2}+lE^{3}}{1 - lE} \right)^{1/2}  \left[ \frac{(2E + 3lE^{2})(1-lE) + l(E^{2}+lE^{3})}{(1-lE)^{2}}   \right]\,\mathrm{ln}\left( 1-e^{-\beta E}\right) \right. \nonumber\\
 & \left. +  E \left(  \frac{E^{2}+lE^{3}}{1 - lE} \right)^{1/2}  \left[ \frac{(2E + 3lE^{2})(1-lE) + l(E^{2}+lE^{3})}{(1-lE)^{2}}   \right]\frac{ \,e^{-\beta E}}{1-e^{-\beta E}}\right\} \mathrm{d}E, 
 \label{entropy2}
\end{split}
\fe
\begin{eqnarray}
\Bar{C}_{V}(\beta,l) &=& \frac{ k_{B} \beta^{2}}{ \pi^{2}}\int_0^{\infty} dE\times\\
&\times& \left\{ E^{2} \left(  \frac{E^{2}+lE^{3}}{1 - lE} \right)^{1/2}  \left[ \frac{(2E + 3lE^{2})(1-lE) + l(E^{2}+lE^{3})}{(1-lE)^{2}}   \right] \frac{ e^{-2 \beta E}}{\left(1- e^{-\beta E}\right)^{2}} \right. \nonumber\\
&+& \left.   E^{2} \left(  \frac{E^{2}+lE^{3}}{1 - lE} \right)^{1/2}  \left[ \frac{(2E + 3lE^{2})(1-lE) + l(E^{2}+lE^{3})}{(1-lE)^{2}}   \right]\frac{e^{-\beta E}}{1-e^{-\beta E}}\right\}. \nonumber
\label{heatcapacity2}
\end{eqnarray}
Initially, we provide the analysis of the Helmholtz free energy displayed in Eq. (\ref{helmontz2}); it is considered within three different scenarios of the Universe: CMB, primordial electroweak epoch, and inflationary era. All these results are demonstrated in Fig. \ref{fs2}, which displays a trivial contribution to the first case, and a decreasing characteristic for the latter two ones. Next, we examined the entropy which was shown in Eq. (\ref{entropy2}); we investigated such thermal function in the same different scenarios of the Universe. All these considerations were shown in Fig. \ref{entropies2}, which exhibited the same trivial contribution to the first case, despite of showing an increase characteristic to the latter two ones. Finally, we studied the heat capacity exhibited in Eq. (\ref{heatcapacity2}); we also examined this thermal function in the same different scenarios of the Universe. All these considerations were shown in Fig. \ref{heatcapacities2}, exhibiting a trivial contribution to the first case as well, and increasing curves for the next two ones.

\begin{figure}[ht]
\centering
\includegraphics[width=8cm,height=5cm]{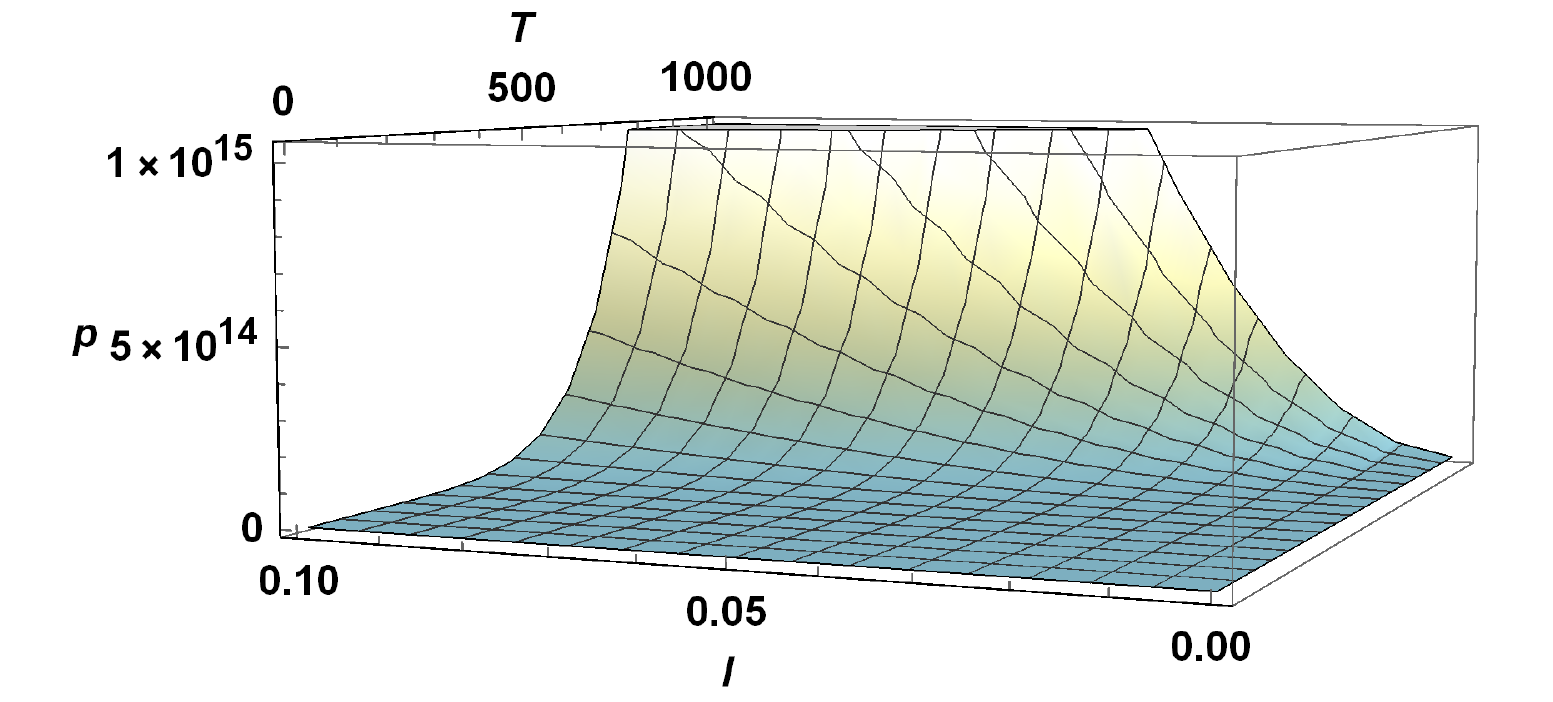}
\includegraphics[width=8cm,height=5cm]{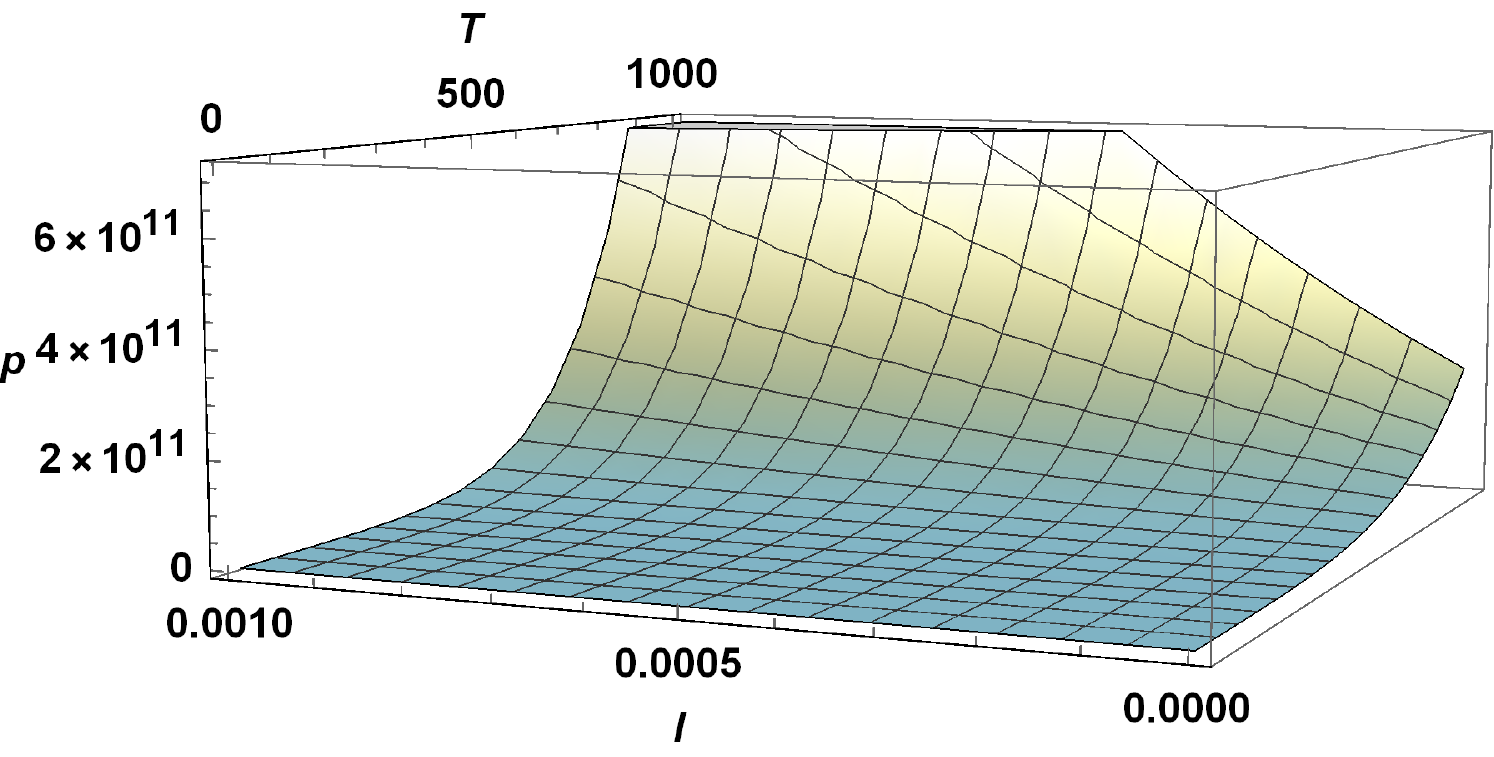}
\includegraphics[width=8cm,height=5cm]{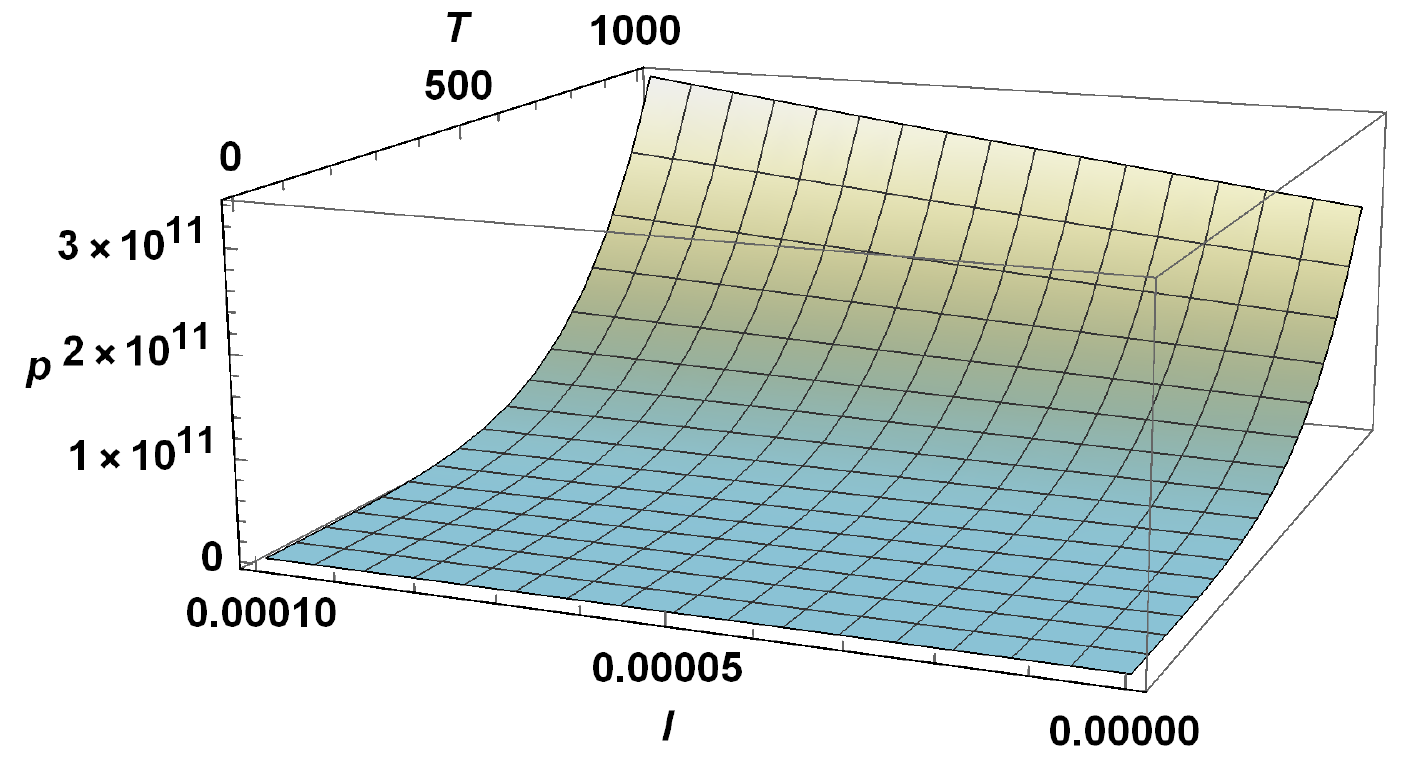}
\includegraphics[width=8cm,height=5cm]{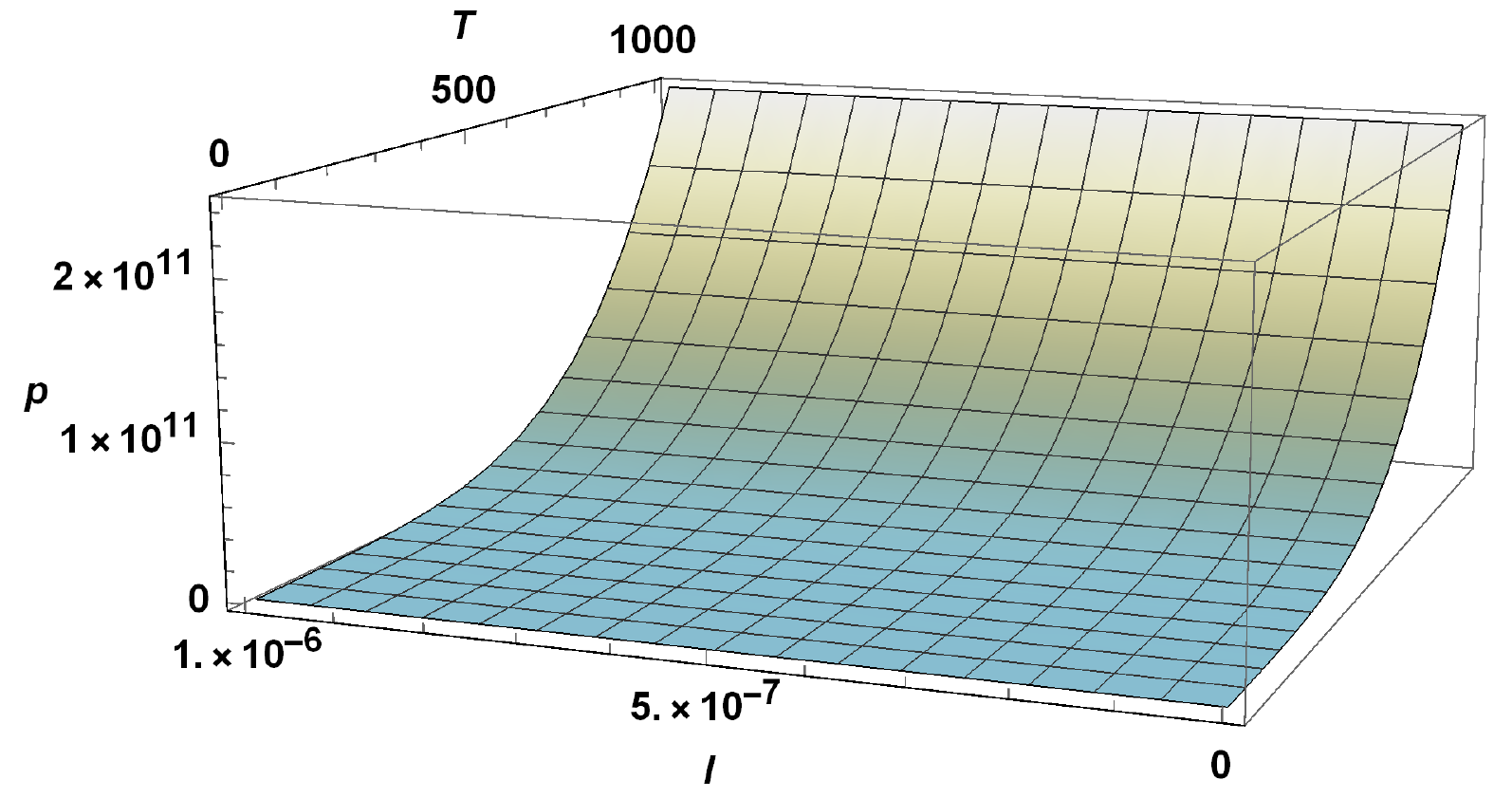}
\caption{The figure shows the equation of states for different values of $p$, $T$, and $l (\text{whose dimension is}\,\, m\cdot kg^{-1/2}\cdot s^{-1})$.}
\label{equationstate2}
\end{figure}

Here, just as we did in the previous section, we also present the analysis of the equation of state. Moreover, since there is no analytical solution to perform our analysis, we have to consider a particular limit to obtain its magnitude as we did before. The limit that we study is $(E^{2} +l E^{3})^{1/2}/(1-l E)^{2} \ll 1$. With it, we obtain the following expression
\ie
p = \frac{1}{45 \pi ^2} T^{6}\left[ 270 l^2 \zeta (5)+4 \pi^4 \frac{l}{T} + 90 \frac{1}{T^{2}} \zeta (3)  \right]. \label{equationofstate2}
\fe
The behavior of Eq. (\ref{equationofstate2}) is displayed in Fig \ref{equationstate2}. Differently with what happens to the model involving $\sigma$, the dispersion relation coming from Eq. (\ref{disper}) has a fascinating feature: the shape exhibited to the equation of states turns out to be sensitive to the modification of the values of $l$. Furthermore, note that, if we consider $l \rightarrow 0$, we obtain
\ie
p = \frac{90\zeta(3)}{45 \pi^{2}} T^{4}.
\fe

\section{Conclusion\label{conclusion}}

We considered the thermodynamical aspects of two higher-derivative Lorentz-breaking theories, namely, CPT-even and CPT-odd ones. Within our study, we focused on the dispersion relations rather than the specific form of the Lagrangians. In this way, we expect that our results can be applied not only to scalar field models as we assumed, but also to specific gauge or spinor field theories. 

We calculated the modification to the black body radiation spectra and to the \textit{Stefan–Boltzmann} law due to the parameters $\sigma$ and $l$. For these theories, we explicitly obtained the corresponding equation of states and the thermodynamic functions as well, i.e., the mean energy, the Helmholtz free energy, the entropy, and the heat capacity. 
Moreover, all the calculations presented in this work were performed taking into account three different scenarios of the Universe: cosmic microwave background, electroweak epoch, and inflationary era.

Furthermore, since the heat capacity rapidly increased with the Lorentz-breaking parameters at high temperatures, perhaps one might expect a phase transition in those scenarios. Nevertheless, a further investigation in this direction might be accomplished in order to provide a proper examination. Thereby, as a further perspective, we suggest the detailed study of the whole field of Horava-Lifshitz gravity model and the possible phase transitions in our models. Another feasible continuation of this study can consist in its application to other higher-derivative Lagrangians of certain known field theories, f.e. those ones discussed in \cite{mariz2019quantum}.


\section*{Acknowledgments}
\hspace{0.5cm}The authors would like to thank the Conselho Nacional de Desenvolvimento Cient\'{\i}fico e Tecnol\'{o}gico (CNPq) for financial support and L. L. Mesquita for the careful reading
of this manuscript and for the suggestions given to us. The work by A. Yu. P. has been partially supported by the CNPq project No. 301562/2019-9. Particularly, A. A. Araújo Filho acknowledges the Facultad de Física - Universitat de València and Gonzalo J. Olmo for the kind hospitality when this work was made. Moreover, A. A. Araújo Filho has been partially supported by Conselho Nacional de Desenvolvimento Cient\'{\i}fico e Tecnol\'{o}gico (CNPq) - 142412/2018-0, and CAPES-PRINT (PRINT - PROGRAMA INSTITUCIONAL DE INTERNACIONALIZAÇÃO) - 88887.508184/2020-00.

\bibliographystyle{ieeetr}
\bibliography{main}

\end{document}